\documentclass[a4paper,usenatbib]{mnras}

\usepackage{footnote}
\usepackage{amsmath}
\usepackage{amssymb}
\usepackage{xcolor}
\usepackage{graphicx}
\usepackage{pdflscape}
\usepackage{url}

\DeclareGraphicsExtensions{.pdf,.png,.jpg}

\newcommand{\OIII}{\ensuremath{\rm [O\,III]}}
\newcommand{\SB}{\ensuremath{\rm erg\,s^{-1}cm^{-2}arcsec^{-2}}}

\title{The spatial extension of extended narrow line regions in MaNGA AGN}

\author[J. Chen et al.]{
  Jianhang Chen$^{1,2}$,
  Yong Shi$^{1,2}$ \thanks{Email: yong@nju.edu.cn},
  Ross Dempsey$^{3}$,
  David R. Law$^{4}$,
  Yanmei Chen$^{1,2}$, 
  \newauthor
  Renbin Yan$^{5}$,
  Longji Bing$^{1,2}$,
  Sandro B. Rembold$^{6,7}$,
  Songlin Li$^{1,2}$,
  Xiaoling Yu$^{1,2}$,
  \newauthor
  Rogemar A. Riffel$^{3,6,7}$,
  Joe R. Brownstein$^{8}$,
  Rogério Riffel$^{7,9}$
  \\
  $^{1}$School of Astronomy and Space Science, Nanjing University, Nanjing 210093, China \\
  $^{2}$Key Laboratory of Modern Astronomy and Astrophysics (Nanjing University), Ministry of Education, Nanjing 210093, China \\
  $^{3}$Department of Physics and Astronomy, Johns Hopkins University, 3400 North Charles Street, Baltimore, MD 21218, USA \\
  $^{4}$Space Telescope Science Institute, 3700 San Martin Drive, Baltimore, MD 21218, USA \\
  $^{5}$Department of Physics and Astronomy, University of Kentucky, 505 Rose Street, Lexington, KY 40506, USA \\
  $^{6}$Departamento de F\'isica, CCNE, Universidade Federal de Santa Maria, 97105-900, Santa Maria, RS, Brazil\\
  $^{7}$Laborat\'orio Interinstitucional de e-Astronomia - LIneA, Rua Gal. Jos\'e Cristino 77, Rio de Janeiro, RJ - 20921-400, Brazil \\
  $^{8}$Department of Physics and Astronomy, University of Utah, 115 S. 1400 E., Salt Lake City, UT 84112, USA \\
  $^{9}$Departamento de Astronomia, IF, Universidade Federal do Rio Grande do Sul, CP 15051, 91501-970, Porto Alegre, RS, Brazil\\
}

\date{Accepted 2019 August 2. Received 2019 July 18}        
\pubyear{2019}                                  

\begin{document}
\label{firstpage}
\pagerange{\pageref{firstpage}--\pageref{lastpage}}
\maketitle

\begin{abstract}
  In this work, we revisit the size-luminosity relation of the extended narrow line regions (ENLRs) using a large sample of nearby active galactic nuclei (AGN) from the Mapping Nearby Galaxies at Apache Point Observatory (MaNGA) survey. 
The ENLRs ionized by the AGN are identified through the spatially resolved BPT diagram, which results in a sample of 152 AGN.
By combining our AGN with the literature high-luminosity quasars, we found a tight log-linear relation between the size of the ENLR and the AGN \OIII ${\rm \lambda 5007}$\AA{} luminosity over four orders of magnitude of the \OIII{} luminosity.
The slope of this relation is $0.42\pm0.02$ which can be explained in terms of a distribution of clouds photoionized by the AGN.
This relation also indicates the AGN have the potential to ionize and heat the gas clouds at a large distance from the nuclei without the aids of outflows and jets for the low-luminosity Seyferts.
\footnotemark[2]
\end{abstract}


\footnotetext[2]{The code used for this paper is available online \url{https://github.com/cjhang/ENLR}}

\section{Introduction}
It is now generally believed that the super-massive black holes at the centers of galaxies co-evolve with their hosts \citep{Hopkins2006, Ho2008, Fabian2012, Kormendy2013, Heckman2014}. 
An AGN can exert feedbacks on its host galaxy and impact its growth and evolution. 
This can occur via radiative processes, in which energetic photons from the AGN photoionize and heat gas in the galaxy, or via mechanical processes such as outflows and jets \citep{Osterbrock2006_Book, King2015}. 
Understanding the nature of the feedback is important for improving our knowledge of galaxy formation and evolution.

Based on the unified model of \citet{Antonucci1993}, the narrow line region (NLR) is an important and ubiquitous component of AGN.
It occurs in a bi-cone away from the obscuring torus, in which the central source illuminates the gas of the host galaxy.
In NLRs, the gas density derived from emission line ratios like ${\rm [S\,II]\lambda 6716/\lambda 6731}$\AA{} and ${\rm [O\,II]\lambda 3729/\lambda 3726}$\AA{} 
is sufficiently low (between $10^2$ and $10^3$ cm$^{-3}$) that its emission is dominated by the forbidden-line transitions;
the gas temperature calculated from the ratio of ${\rm [O\,III] (\lambda 4959 + \lambda 5007)/\lambda 4363}$\AA{} is around the photoionization balance temperature of $10^4$K;
the ionization parameter defined as the ratio between the photon density and the electron density has been mostly set as $U \sim 0.01$ with only 0.5 dex of variation in observations \citep{Bradley2004, Nesvadba2008, Osterbrock2006_Book}.
The size of a NLR is first thought to be sub-kpc scales, but further observations using narrow band images and long-slit spectra revealed some extended ionized nebulae up to several kpc or even tens of kpc in some galaxies \citep{Heckman1981, McCarthy1987, Keel2012, Liu2013, Obied2016}. 
These extended nebulae are called the ``extended emission line regions'' (EELRs) if they are formed by stellar processes, or ``extended narrow line regions''(ENLRs) if they are mainly produced by the AGN activity, but the two terms are sometimes mixed in use \citep{Stockton1987, Unger1987, Husemann2013}.
The ENLR got its name from luminous radio galaxies \citep{Unger1987} and its formation was thought to be the interaction between the jet and its ambient gas \citep{Heckman1981, Boroson1985, Stockton1987, Fu2009}. 
But further studies have found that many Serfert galaxies also show similar ENLRs \citep{Bennert2002, Schmitt2003, Schmitt2003b}.

Theoretically, many models have been proposed to explain the formation of (E)NLR.
The standard photoionization models assume a set of constant density clouds ionized by a power-law or broken power-law ionizing source \citep{Osterbrock2006_Book}, and shock excitation is added for some ENLRs associated with jets \citep{Dopita1995, Dopita1996, Solorzano2001}.
However, the standard photon-ionization models have several problems.
The observed strong coronal lines in some Seyfert galaxies cannot be produced in these models \citep{Dopita2002, Groves2004}, and the relationship between the ENLR size and the AGN luminosity has a slope that differs from the predicted value of 0.5 \citep{Schmitt2003b, Netzer2004}.
More recent ENLR models replace the constant density with multicomponent gas densities or include the dust contribution \citep{Dopita2002, Groves2004, Groves2004b, Dempsey2018} which produce different slopes of the size-luminosity relation.
A better constraint on the slope from the observation is required to distinguish different models.

However, the derived slopes are far from consistent with each other in observations.
The observation done by \citet{Bennert2002} found $R_{\rm NLR} \sim L_{\rm [O\,III]}^{0.52 \pm 0.06}$ that is close to the prediction of the standard photon-ionization model, while \citet{Schmitt2003b} found a much flatter relation with a slope $\sim 0.33$.
Since then, many works have revisited this relation with new observations \citep{Greene2011, Hainline2013, Liu2014, Husemann2014, Bae2017, Sun2018, Fischer2018}, but their derived slopes range from 0.23 to 0.52.
The difference may be attributed to their different definitions of the ENLR size, different proxies of the AGN luminosity or different sensitivities. 
Additionally, most of these works focused on the high-luminosity AGN with a relatively small dynamic range in the AGN luminosity, which limits the accuracy of the derived slope. 
The advent of integral field unit (IFU) spectroscopy, especially the massive IFU surveys like SDSS-IV/MaNGA \citep{Bundy2015}, offers a new opportunity to measure the size-luminosity relation of ENLRs with a much larger and uniform sample.
The spatially resolved spectra also enable us to identify the ENLR by the emission line diagnostics, which is an effective way to isolate the ENLR for low-luminosity AGN.
With the large sample available, we can get a better constraint on the slope of size-luminosity relationship of ENLRs in a large dynamic range of AGN luminosity, which can be helpful to answer the formation of ENLRs and to understand the feedback of AGN.

The paper is organized as follows. In section 2, we describe the basic information of MaNGA and sample selection. The methods used to measure the strength of AGN and determine the size of ENLR are described in section 3. The size-luminosity relation of ENLR is described in section 4. Models and possible mechanisms that contribute to the extension of ENLR are discussed in section 5. Finally, we summarize our work in section 6. A flat ${\rm \Lambda}$CDM cosmology with ${\rm \Omega_\Lambda}$=0.7, $\Omega_M$=0.3 and ${\rm H_0=70\,km\,s^{-1}Mpc^{-1}}$ is assumed throughout this work.

\section{Data}
\subsection{MaNGA overview}
As one of the three major programs of Sloan Digital Sky Survey IV (SDSS-IV), MaNGA \citep{Bundy2015, Blanton2017} uses the 2.5 m Sloan Foundation Telescope \citep{Gunn2006}, aiming at obtaining IFU observations of over 10,000 nearby galaxies from 2014 to 2020 \citep{Law2015, Drory2015, Yan2016, Yan2016b}. 
Currently, more than half of the sample has been observed, including 6430 unique targets, some of them have been published as Data Release 15 (DR15) of SDSS-IV \citep{Aguado2019}. 
All the analyses in this work are based on the eighth internal MaNGA Product Launches (MPL-8).

Each target of MaNGA was observed with one of the specially designed hexagonal bundles ranging from 19 to 127 fibers \citep{Drory2015}, and the spectra were fed to the two BOSS spectrographs with an overall wavelength coverage from 3600 to 10,300${\text \AA}$ \citep{Smee2013}. 
The typical seeing of Apache Point Observatory is 1.5$''$, but the final spatial resolution of MaNGA is about 2.5$''$ including the smearing from telescope and instruments, which corresponds to 1-2 kpc at the redshift range of $0.01 < z < 0.15$ \citep{Wake2017}.
For the spatial coverage of the bundles, about 30\% of the sample has uniform coverage larger than 2.5 effective radii ($R_e$, the radius containing 50\% of the light of the galaxy) and the rest has at least 1.5 $R_e$. 
All the data have been reduced by the Data Reduction Pipeline (DRP) \citep{Law2016} and analyzed by the Data Analysis Pipeline (DAP) \citep{Westfall2019}. 
Both the software and data are available in the public release \citep{Aguado2019}. 
All those features make MaNGA the ideal sample for selecting the AGN with extended narrow line regions in nearby universe.

\begin{figure*}
  \centering
  \includegraphics[width=0.9\linewidth]{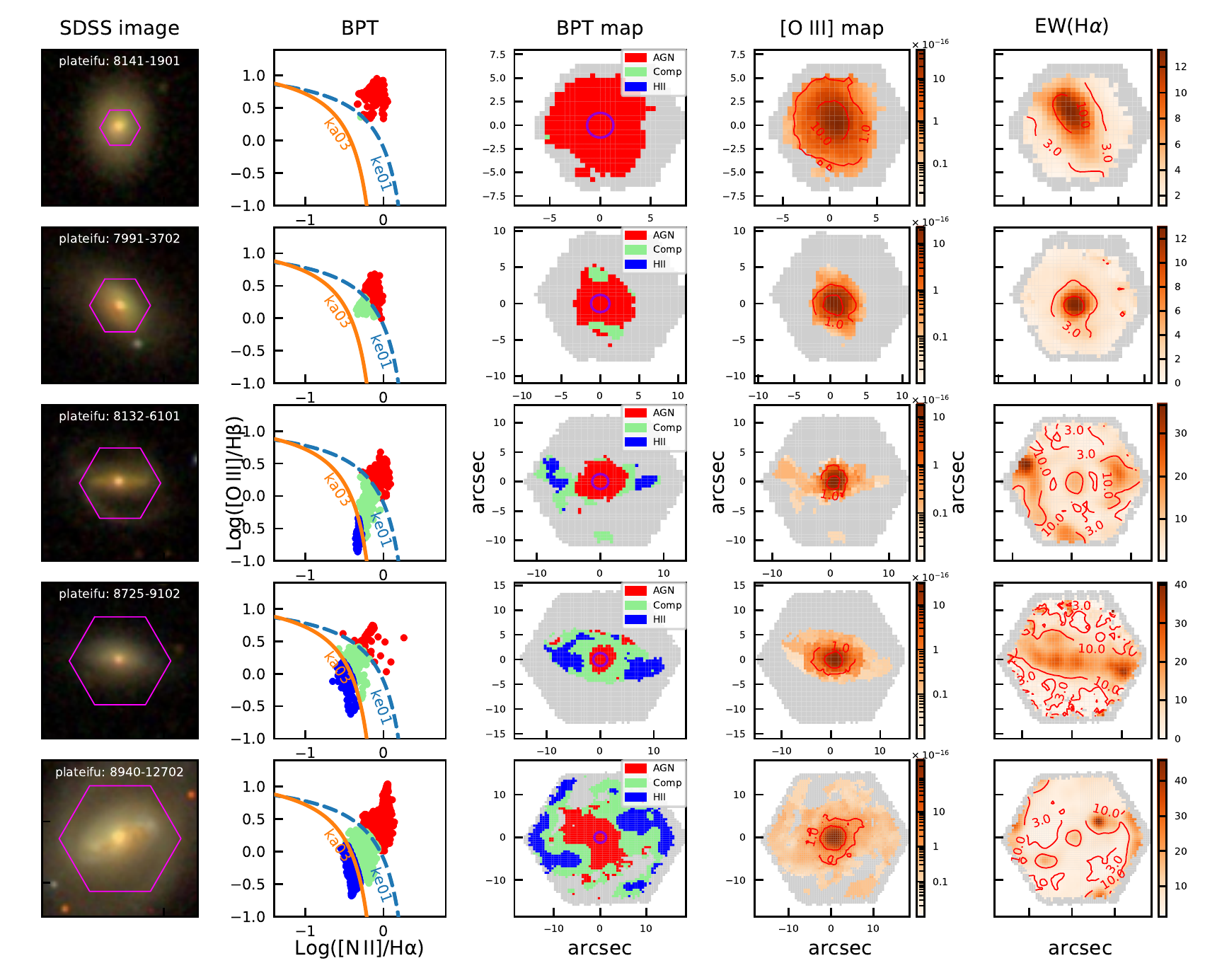}
\caption{
  Five example galaxies observed by different bundles of MaNGA. 
  Their fiber numbers are 19, 37, 61, 91, 127 from top to bottom. 
  The bundle edges are indicated by the magenta hexagon in the SDSS RGB images and the grey shaded region in maps. 
  In the second column, we take each spaxel of the galaxy and plot it on the BPT diagram, using the same separation curves as in Fig. 1. 
  These spaxels and their classifications are shown in projection in the third column, 
  We are most interested in spaxels above the Ke01 line, in which the spectra are dominated by AGN activity \citep{Kewley2001, Kewley2006}. 
  The purple circles in the center of each plot in the third column depict the FWHM of the PSF in the g-band.
  The fourth column are the surface brightness maps of \OIII{} after attenuation correction, the contours show the value of the surface brightness in units of $10^{-16}$\SB{}.
  The last column depicts the maps of EW(H$\alpha$) in angstrom.
}
  \label{fig:detail_sample}
\end{figure*}

\subsection{Sample selection}
The most standard way to select the AGN in the optical bands is through the well known BPT diagram \citep{Baldwin1981, Kauffmann2003, Ho2008}. It uses the emission line ratios ${\rm log [O\,III] \lambda5007/H\beta}$ vs ${\rm log [N\,II] \lambda 6583/H\alpha}$ or ${\rm log [S\,II] \lambda\lambda 6717,31/H\alpha}$ to classify the galaxies as Seyferts, LINERs, star-forming and composite galaxies based on their dominant ionization mechanisms \citep{Baldwin1981, Kewley2001, Kauffmann2003, Kewley2006}.
We mainly focused on the ${\rm [N\,II]}$-based BPT diagram, which gives a better separation between the AGN and ${\rm H\,II}$ region \citep{Belfiore2016}, but using the ${\rm [S\,II]}$-based BPT diagram does not affect the main results.

\begin{figure}
  \centering
  \includegraphics[width=0.9\linewidth]{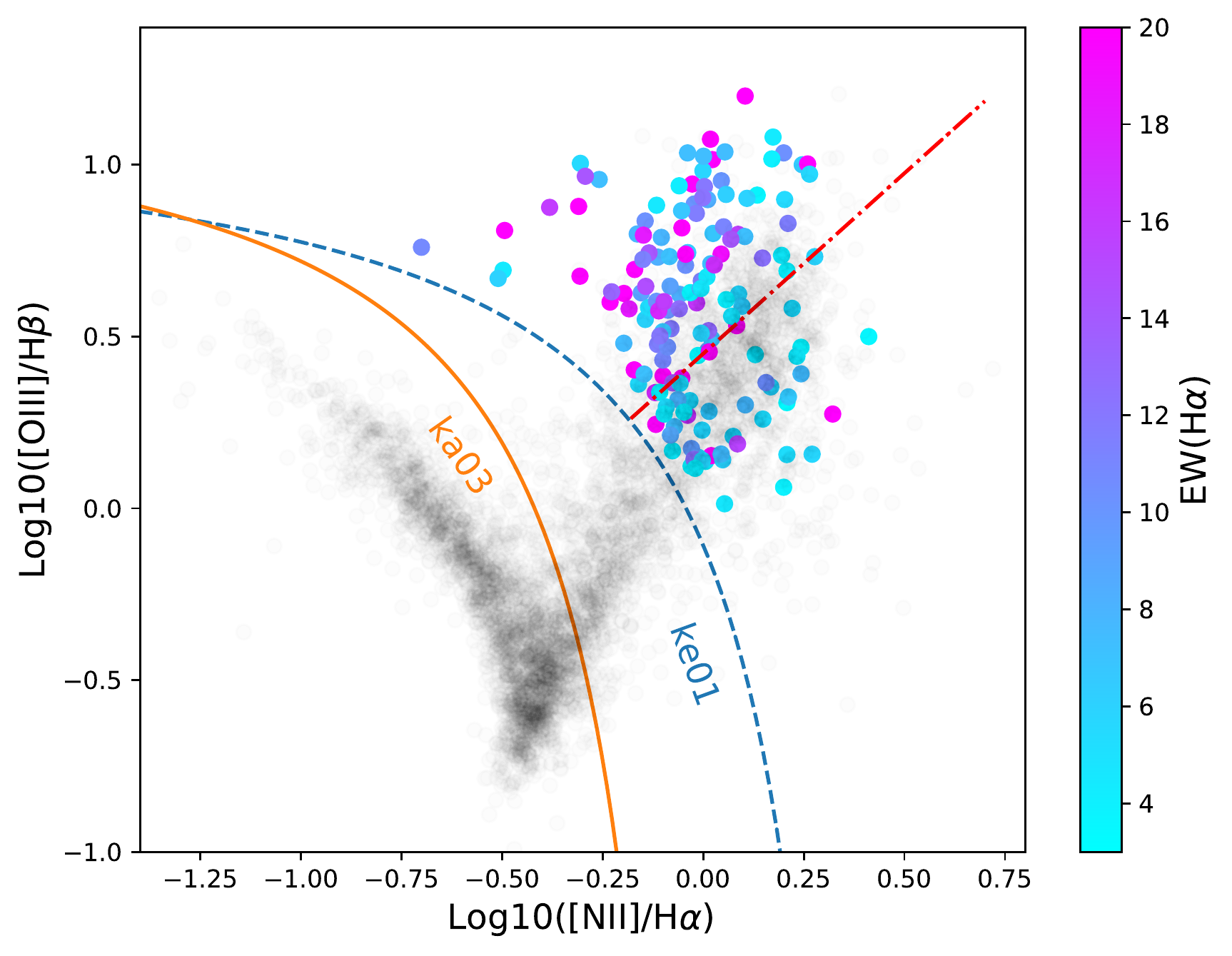}
  \caption{
  All the AGN candidates selected from MaNGA MPL-8.
  All the line ratios derived from the median ratio of each galaxy in their central region(within the radius of 3$''$).
  Light grey circles (background) show all the MPL-8 galaxies and color coded circles are our selected galaxies with their colors standing for the median EW(H$\alpha$) of the central region as before.
  The solid orange line is the empirical line (Ka03) which separates the star-forming galaxies and the composite \citep{Kauffmann2003}, the blue dashed line (Ke01) is the maximum ionizing boundary of star-forming activity \citep{Kewley2001} and the red dot-dashed line is the S07 line \citep{Schawinski2007} which is an empirical separation between Seyferts and LINERs. 
  We did not distinguish between Seyfert and LINER in thiis work, but most of our galaxies are Seyferts according to this line.}
  \label{fig:sample}
\end{figure}

Our sample selection method is adapted from \citet{Cid_Fernandes2010} and \citet{Rembold2017}.
We first plotted each spaxel within the central radius of 3$''$ of each MaNGA target on the BPT diagram, and then selected those galaxies with more than 2/3 spaxels classified as AGN.
For each galaxy, we only used the spaxels with at least 3 $\sigma$ detection of all the emission lines needed for the BPT diagram.
Additionally, we required the equivalent width (EW) of H$\alpha$ to be larger than 3${\text \AA}$ in order to reduce the contaminations from the diffused ionized gas (DIG) \citep{Cid_Fernandes2010, Lacerda2018} (also see some discussion in \citet{Zhang2017}). 
Fig. \ref{fig:detail_sample} illustrates five AGN examples along with the BPT maps and EW of H$\alpha$ maps. 
Besides DIG, \citet{Belfiore2016} also suggested the central and extended low-ionization regions (LIERs) are more likely to be ionized by diffused stellar sources.
Based on their classification, several of our LINER-like AGN belong to the cLIER (central LIER), but since their EW(H$\alpha$) are larger than 3${\text \AA}$ and unlikely contaminated by stellar radiation, we still included them in our sample.
With this method, after removing 16 close merging pairs, we finally found 152 AGN candidates from MPL-8.
All the candidates are shown in Fig. \ref{fig:sample} and listed in Tab. \ref{tab:1}.
Both type-I and type-II AGN are selected, but the majority of them are type-II Seyferts.
Their average uncorrected \OIII ${\rm \lambda 5007}$\AA{} (hereafter \OIII{}) luminosity is about ${\rm 10^{40.5} erg\,s^{-1}}$, who mostly belong to low-luminosity Seyferts. 

\section{Methods}
\subsection{Spectrum fitting}
The DAP products of MaNGA already provide the full spectrum fitting for each galaxy.
The technical details of their ``hybrid'' approach can be found in \citet{Westfall2019}. 
In brief, three fitting iterations are applied to get the emission lines of each spaxel. 
For the first iteration, the continuum is binned to reach a global g-band S/N of 10 to get the binned stellar kinematics. 
For the second iteration, the stellar kinematics are kept fixed to model the continuum and emission lines simultaneously in each spaxel. 
All the emission lines are also modeled with the same kinematics to provide the initial starting guess for the next iteration.
Finally, the velocity dispersions of different emission lines are fitted independently but the velocities are constrained to be the same in each spaxel. 
This fitting results are given in the publicly released products of DR15 \citep{Belfiore2019}.

\begin{figure*}
  \centering
  \includegraphics[width=0.9\linewidth]{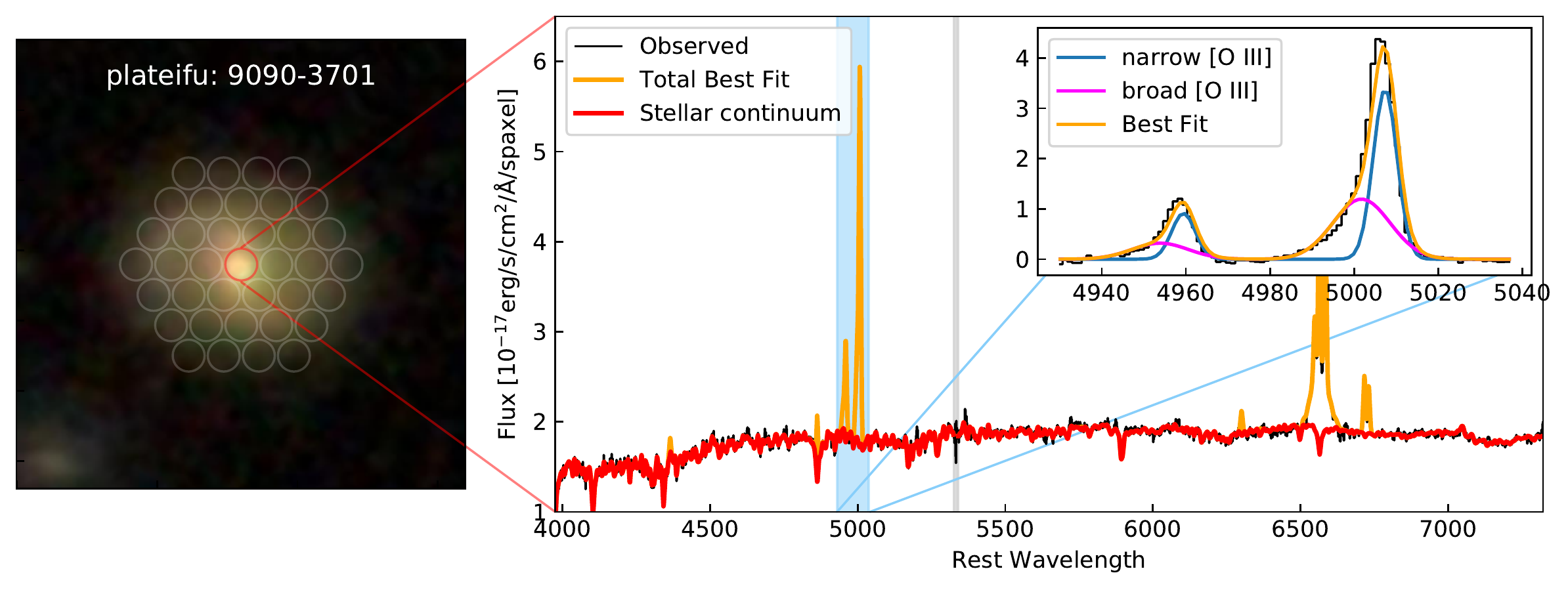}
\caption{
  Example of full spectrum fitting with broad emission lines.
  On the left, it is the SDSS image with fibers superimposed. 
  On the right, it is a spectrum extracted from one of the spaxels belong to the red fiber.
  The black line shows the observed flux in this spaxel, with the sky lines masked by grey shaded area. 
  The stellar continuum (red) is from DAP hybrid fitting results \citep{Westfall2019} and the total best fit (orange) is the stellar continuum pluses the fitted emission lines.
  The zoom-in panel in the right show the emission line fitting results around \OIII{}(the light blue shaded area), with the narrow components in blue and broad components in magenta. 
  The stellar continuum has been subtracted when we fit the emission lines.
}
  \label{fig:spectrum_fitting}
\end{figure*}

For the AGN candidates we are interested in, the emission lines generally cannot be modeled well with the single Gaussian profiles used in DR15. 
We thus fitted the emission lines by adding additional broad components to the H$\alpha$ and H$\beta$ lines with velocity dispersions larger than 800 km/s to represent the possible signals from broad line regions. 
To trace possible strong outflow from our AGN, a broad component with velocity dispersions larger than 600 km/s is added to the \OIII{} doublet during the fitting \citep{Harrison2014}.
An example of the full spectrum fitting with broad emission lines is illustrated in Fig. \ref{fig:spectrum_fitting}.
We only accepted the additional broad component when the improvement in the fitting is statistically significant at the 3$\sigma$ level and pass the F-test.

\subsection{The AGN bolometric luminosity}
The luminosity of \OIII{} is often used as a proxy for the AGN bolometric luminosity \citep{Heckman2004, Kauffmann2009, Heckman2014}.
It has been widely used to study the properties of the ENLRs \citep{Husemann2008, Husemann2014, Sun2018}.
To remove star-formation contamination to the AGN \OIII{} luminosity, we only integrated the emission of AGN-dominated spaxels as identified by the BPT as the total AGN \OIII{} luminosity.
This process is illustrated in the second and third columns of Fig. \ref{fig:detail_sample}, where the red spaxels of the IFU are dominated by AGN ionization according to Ke01 line \citep{Kewley2001}.
The most controversial spaxels are the ones between the Ke01 and the Ka03 lines, which can either be produced by shock excitation or the mixture of AGN and star formation activities.
But their contribution to the total \OIII{} luminosity is small compared to that of the AGN regions, so the \OIII{} from the composite region is treated as an additional error of the overall \OIII{} luminosity. 

The dust attenuation is corrected by the Balmer decrement and the dust reddening curve of \citet{Calzetti2001}, with a assumption of ${\rm A_V / E(B-V) = 3.1}$ and intrinsic ${\rm H\alpha /H\beta}$ ratio of 3.1 \citep{Kewley2006}.
The flux maps of H$\alpha$ and H$\beta$ are rebinned by \emph{VorBin} \citep{Cappellari2003} to reach a global S/N of at least 10, and the sigma-clipping of 5 is applied to the derived E(B-V) maps to mask anomalous values. 
All the remaining analyses are based on the dust corrected \OIII{} maps.

\subsection{The sizes of the extended narrow line regions}
\begin{figure*}
\centering
\begin{minipage}{0.45\textwidth}
  \centering
  \includegraphics[width=.9\linewidth]{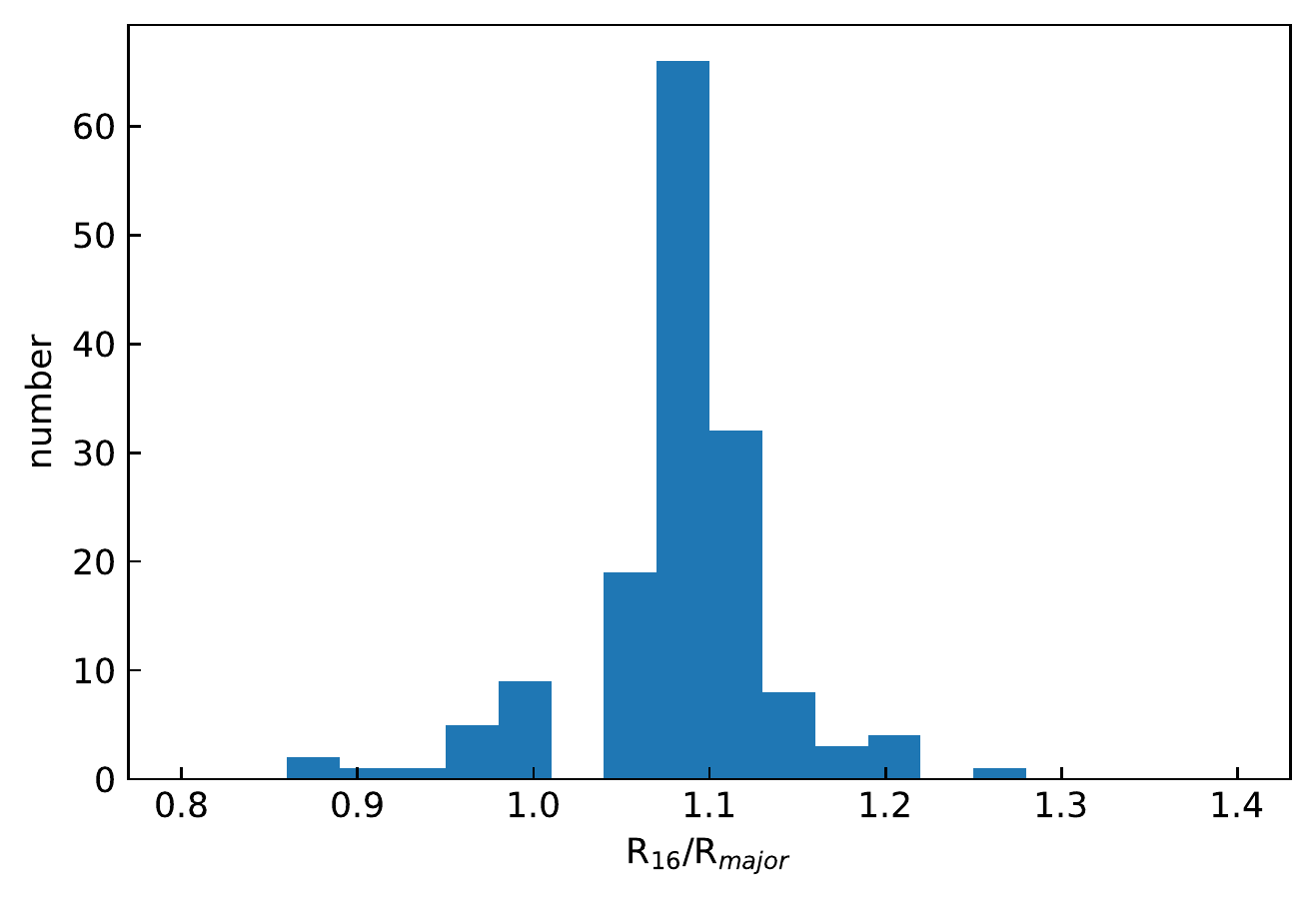}
\end{minipage}%
\begin{minipage}{0.45\textwidth}
  \centering
  \includegraphics[width=.9\linewidth]{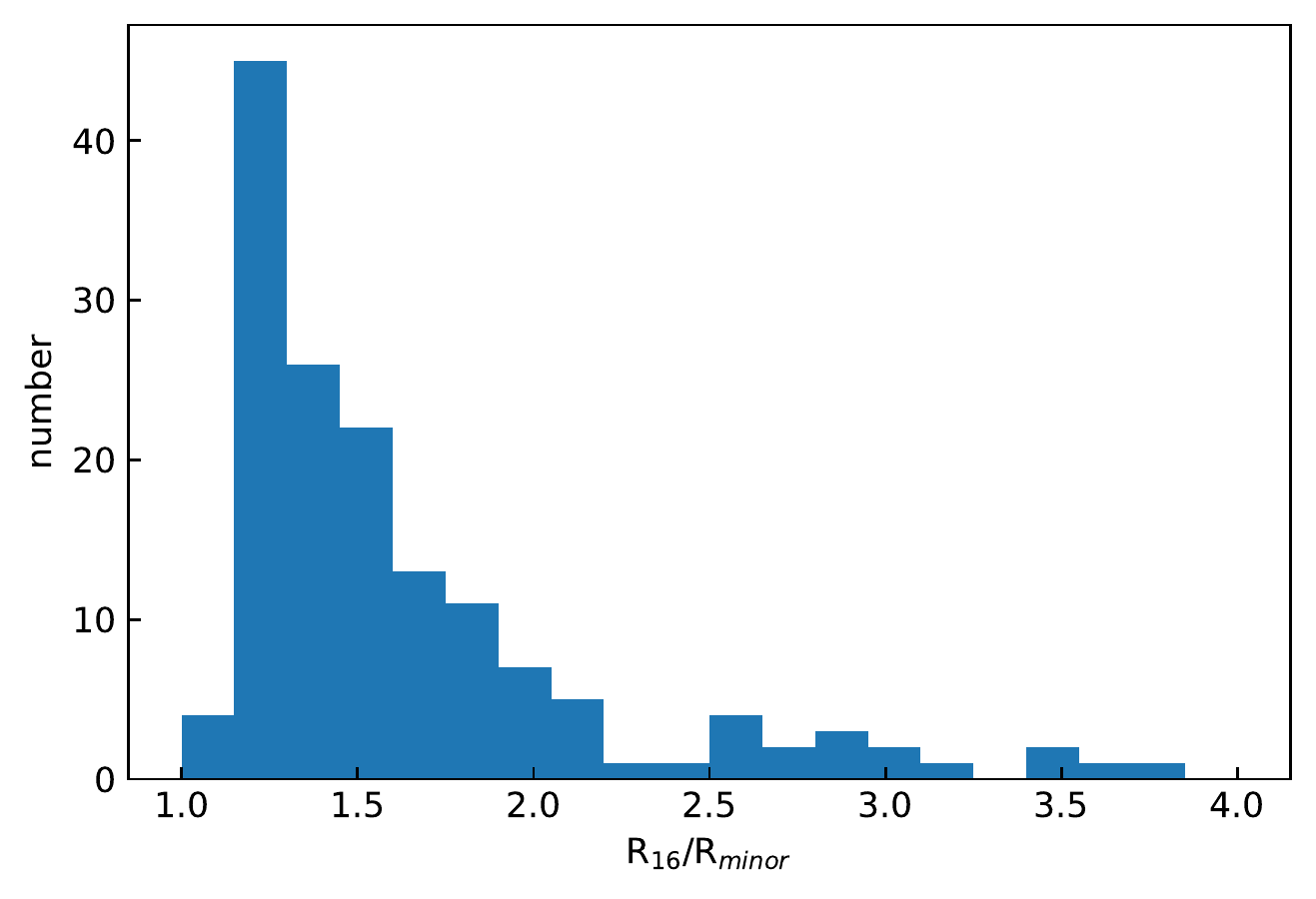}
\end{minipage}
\caption{
Comparing size measurements based on IFU and mock long-slit observations.
Two directions of the long-slit are applied, one aligned with the major axis (left) and the other with the minor axis (right) of the galaxies. 
The X axis is the ratio between two size at the surface brightness of $10^{-16}$\SB{}.
}
\label{fig:mock_longslit}
\end{figure*}

The size of ENLRs have been defined in different ways in the literature.
\citet{Bennert2002} and \citet{Schmitt2003b} used the Hubble Space Telescope (HST) to obtain narrow band images of \OIII{}, and adopted the maximum 3$\sigma$ detected radius as the radius of the ENLR. 
This method is subject to the instrumental sensitivity limit that could be very different in different observations.
Studies with long-slit spectroscopic observations define the radius based on isophote \citep{Greene2011, Hainline2013, Hainline2014}, or the distance at which the ionization state changes from AGN to star-forming activities \citep{Bennert2006a, Bennert2006b}.
The long-slit based observations also have drawbacks: the morphology of ENLR is sometimes irregular so that the derived size depends on the orientation of slits \citep{Greene2011, Husemann2013}.
We have compared the measured size based on the IFU and the mock long-slit observation in Fig. \ref{fig:mock_longslit} following the method discussed below. In most cases, long-slit observations tend to underestimate the true size of ENLR.
IFU spectroscopic data allow us to use two-dimensional maps to define the sizes of ENLRs. 
Common definitions include the radius of a specified \OIII{} surface brightness isophote \citep{Liu2013, Liu2014}, or the \OIII{} flux weighted radius \citep{Husemann2013, Husemann2014, Bae2017}.
We followed the same method as \citet{Liu2013} but chose a different threshold.
The isophote threshold of $10^{-15}$\SB{} was used for quasars related studies. 
This is suitable for such bright objects but are not as useful for fainter Syferts in the our sample as it will leave a large number of AGN undetected.
The typical 3 $\sigma$ depth of the MaNGA observation in \OIII{} surface brightness can reach $10^{-17}$\SB{}.
For our AGN sample, the majority of AGN spaxels have surface brightnesses above $10^{-16}$\SB{} which is thus adopted in this work as the threshold to define the sizes of the ENLRs (hereafter $R_{16}$).
If all spaxels are above this threshold, we extrapolated the fitted \OIII{} surface brightness profile to determine $R_{16}$ (see Sec. 3.4 for more detail).
It should be noted that the surface brightness can be affected by cosmological dimming, which has a scale factor of $(1+z)^4$ \citep{Liu2013, Hainline2014}.
That is important for works trying to compare sample with different redshift, especially for high redshift quasars.

\subsection{Effects of PSF}
The spatial properties of MaNGA galaxies are smeared by the seeing from the atmosphere, the telescope and the instruments. The final PSF of each band has been reconstructed by the DRP \citep{Law2016}.
To reduce the PSF effects, we first derived the 1-D surface brightness profile of the \OIII{} maps of AGN spaxels using the Ellipse fitting provided by \emph{Astropy} affiliated package \emph{photutils} \citep{AstroPy, photutils}.
The profile is then fitted by the S\'ersic light distribution:
\begin{equation}
  \ln I(r) = \ln I_0 - k r^{1/n},
  \label{eq:sersic}
\end{equation}
where the S\'ersic index $n$ is kept in the range [0.5, 10] and $k$ in the range [0.01, 10].

\begin{figure*}
  \centering
  \includegraphics[width=0.9\linewidth]{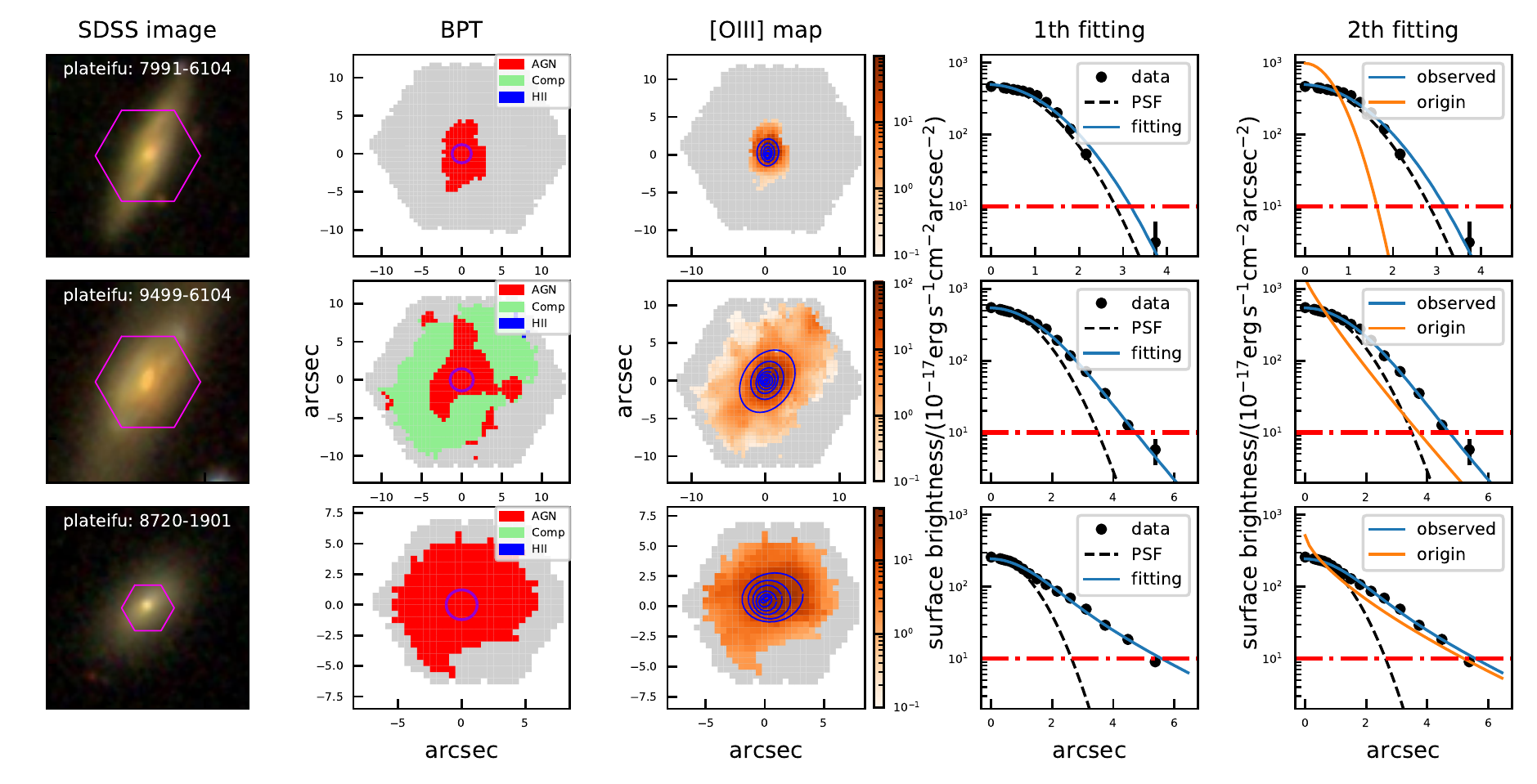}
  \caption{
    The effects of PSF on the distribution of \OIII{} surface brightness profile. 
    The first two columns are the same as Fig. \ref{fig:detail_sample}. 
    The third column also shows the flux map of \OIII{} as in Fig. \ref{fig:detail_sample}, but with the fitted isophotes overlaid.
    The last two columns show the two steps for surface brightness fitting. 
    The black dots are the mean surface brightness of each isophote overlaid on the \OIII{} map. 
    The blue solid lines are the fitted S\'ersic profile, the black dashed line shows the Gaussian modeled PSF of g-band and the orange solid lines show the PSF-corrected surface brightness profile. 
    The red dot-dashed lines are the threshold of surface brightness $10^{-16}$\SB{}.
    The surface brightness profile of the first target is significantly altered by the PSF, while the third one is almost immune from the PSF.
  Their size difference between the two fitting at $10^{-16}$\SB{} is 49\%, 19\% and 6\%.}
  \label{fig:psf_demo}
\end{figure*}

Two ways are applied to fit each surface brightness profile. 
In the first one, the profile is fitted with the S\'ersic profle without the PSF, while in the second one, the profile is fitted with the S\'ersic profile convolved with a Gaussian modeled g-band PSF, where the FWHM of PSF is provided by the DRP \citep{Law2016}.
The fitting results of the first way are taken as the observed profile and the results of second way are regarded as the intrinsic profiles after correcting the PSF.
Examples illustrated in Fig. \ref{fig:psf_demo} show three surface brightness profiles with different PSF. 
For the marginally resolved profile (like the first one in Fig \ref{fig:psf_demo}), its intrinsic profile is severely affected by the PSF and the size differs nearly 50\% compared with the observed one.
While for those clearly resolved profile (like the last one), its intrinsic and observed profiles remain almost the same with only 6\% difference.
In most cases, the size derived from the second way is used, but if the fitting iterations of the second way end by reaching the boundaries of parameters, it is labeled as unsuccessful and the radius derived from the first way is used as the upper limit.
All the galaxies with failed fitting in the second way are set as unresolved ENLRs in Tab. \ref{tab:1}. 
If both ways failed, 15 galaxies, they are excluded from the later analysis.
The uncertainties of all the final radii are the standard deviations calculated by 100 Monte Carlo simulations which randomly adding Gaussian noise to the two dimensional flux map and repeating all the steps mentioned above to re-calculate the radii.

\section{The size-luminosity relation of ENLR}
\begin{figure*}
  \centering
  \includegraphics[width=0.9\linewidth]{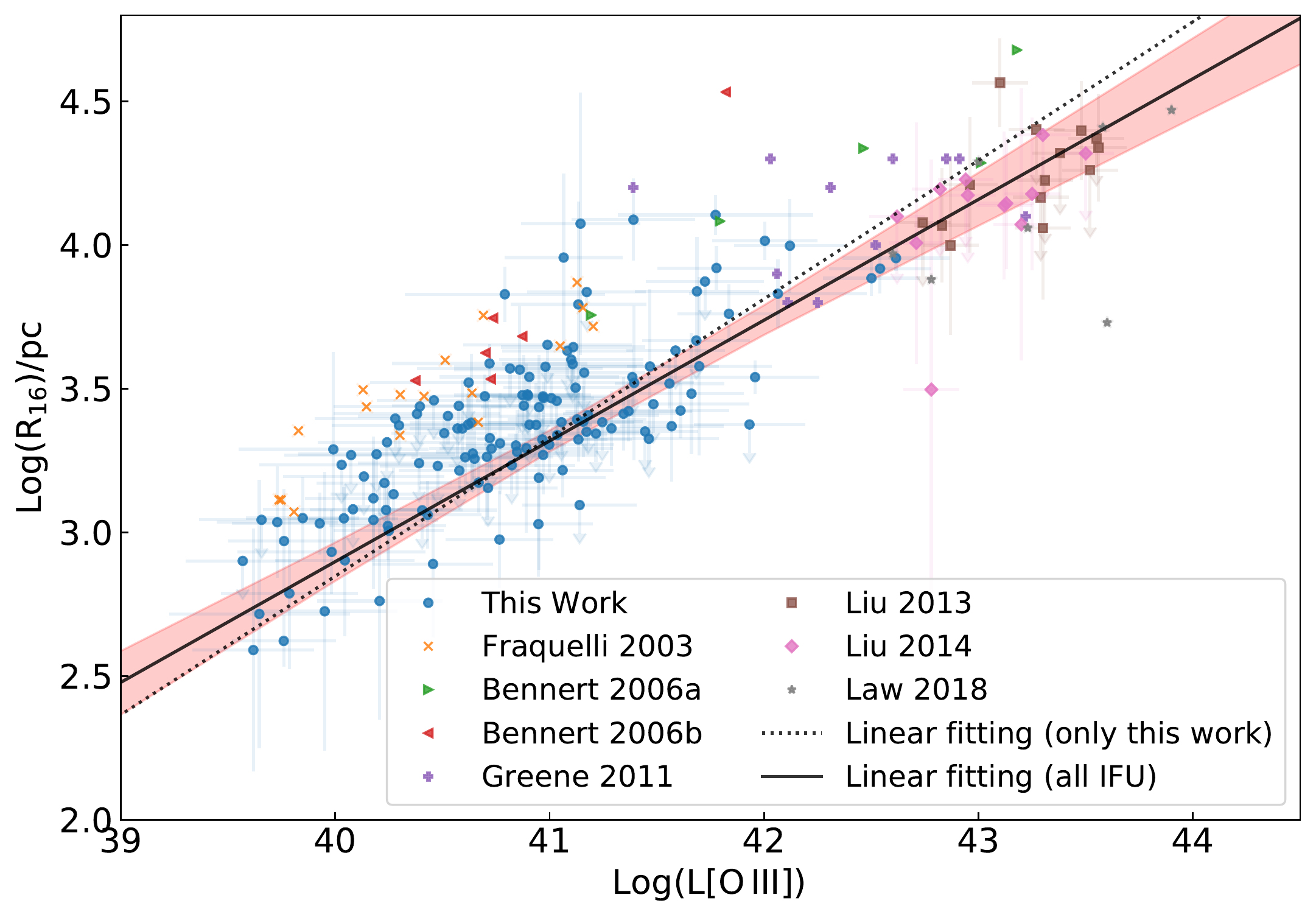}
  \caption{
    The overall relation between the size of ENLR and luminosity of \OIII{}.
    R$_{16}$ is the ENLR size at a surface brightness cut of $10^{-16}$\SB{} and the \OIII{} luminosity is measured from the AGN region of each galaxy with dust extinction corrected. 
    The data from the AGN candidates in this study are shown in blue circles with 1$\sigma$ error bars, the undetected galaxies are labeled by upper limits. 
    We also plot the data measured from Seyfert galaxies \citep{Fraquelli2003, Bennert2006a, Bennert2006b}, luminous quasars \citep{Greene2011, Liu2013, Liu2014} and high redshift faint AGN \citep{Law2018}.
    But only the IFU data from \citet{Liu2013} and \citet{Liu2014} are included in the fitting.
    Our best linear fit results in a slope of 0.42 $\pm$ 0.02 (black solid line), with the 95\% confidence level in light magenta shadow.
    If only our data is used, the derived slope is 0.49$\pm$ 0.04 (black dotted line).
  }
  \label{fig:all_fitting}
\end{figure*}

All the derived data of our sample are summarized in Table 1. Their sizes of ENLRs and luminosities are plotted in Fig. \ref{fig:all_fitting} with 1$\sigma$ error bars.
We also included the quasars data from \citet{Liu2013, Liu2014}, which were obtained by the IFU spectroscopic observations on the Gemini telescope and had reached the surface brightness depth of $10^{-16}$\SB. 
It helps to extend our luminosity range from 3 dex to 4 dex, which is important to constrain the slope of the size-luminosity relation.
We applied the same method on their derived surface brightness profiles to get $R_{16}$, but added an additional 20\% uncertainty to account for possible errors introduced by the fitting processes.
For their \OIII{} luminosity, 20\% error was added for the type-II quasars \citep{Liu2013} and 10\% additional error for type-I quasars \citep{Liu2014} to account for the errors caused by dust extinction.
\citet{Law2018} also provided the $R_{16}$ of their seven IFU observed faint AGN at $z\sim2$, but it needs the extrapolation to our threshold. 
The robust error is hard to estimate for their data, so we did not include them in the final fitting.
In addition, we also included many other results based on long-slit spectra for comparison.
Like the quasars from \citet{Greene2011} and Seyferts from \citet{Fraquelli2003, Bennert2006a, Bennert2006b}. 
All their $R_{16}$ are extrapolated from their power-law fitting results.
Based on all the valid IFU observations, a log-linear fit is derived by using the Bayesian method of \citet{Kelly2007}.
The best-fit solution we have obtained is:
\begin{equation}
\log\left(\frac{R}{\text{pc}}\right) = (0.42 \pm 0.02) \log\left(\frac{L_{\rm [O\,III]}}{\rm erg\,s^{-1}}\right) - (13.97 \pm 0.95)
  \label{eq:final}
\end{equation}
It is shown in Fig. \ref{fig:all_fitting} with 95\% confidence interval.
The slope derived by our data alone is $0.49\pm 0.04$ shown as the dotted line in Fig. \ref{fig:all_fitting}.
This is steeper than some previous studies \citep{Schmitt2003, Greene2011, Liu2014, Hainline2014}, and more close to 0.5 \citep{Bennert2002}.
Since most of the previous results were based on small samples that are limited in the luminosity range, our results should be a better constraint.

\begin{figure}
  \centering
  \includegraphics[width=0.9\linewidth]{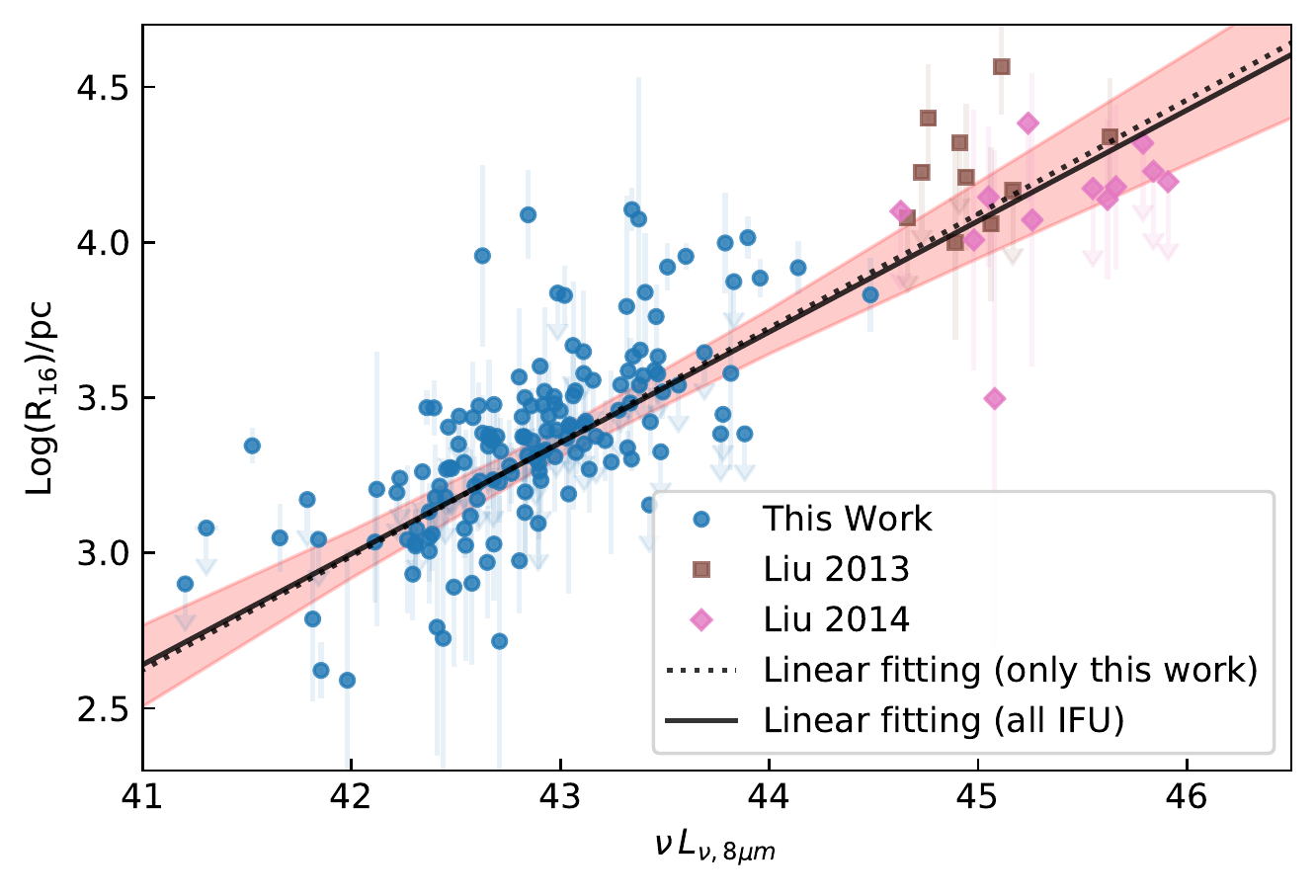}
  \caption{
    Similar to Fig. \ref{fig:all_fitting}, but the luminosities have changed to ${L_{\rm 8 \mu m}}$. The fitting has a slope of $0.36\pm 0.03$ (black solid line) if all the data is used and a slope of $0.37\pm 0.05$ (black dotted line) if only our data is used.
  }
  \label{fig:all_fitting2}
\end{figure}

\citet{Hainline2014} and \citet{Liu2014} argued that their flattened slope might due to the non-linear relation between the ${L_{\rm [O\,III]}}$ and ${L_{\rm bol}}$.
After replacing the ${L_{\rm [O\,III]}}$ with ${L_{\rm 8 \mu m}}$, a slope near 0.44 could be obtained.
In order to check the relation between $R_{16}$ with ${L_{\rm 8 \mu m}}$ based on our sample, we cross matched MaNGA with Wide Field Infrared Explorer (WISE) catalog \citep{Wright2010} and interpolated the ${L_{\rm 8 \mu m}}$ based on photometry at 4.6 $\mu m$ and 12 $\mu m$.
  As shown in Fig. \ref{fig:all_fitting2}, the overall slope is $0.36\pm 0.03$ if including the data from \citet{Liu2013, Liu2014} and the slope changes to $0.37\pm 0.05$ if only our data are used.
  We caution the reliability of the results based on the IR luminosity, as a significant contribution from star formation is expected to contaminate the ${L_{\rm 8 \mu m}}$ for our Seyferts.
  Although the SED fitting can be used to estimate the star-formation contribution to the IR luminosity, but only a tiny fraction of our sample has far infrared photometry which makes it hard to get a reliable correction.

For the more recent works,
\citet{Sun2018} proposed a new method to acquire large samples of AGN with extended \OIII{} maps. 
Their \OIII{} maps were reconstructed from the broad band-images by carefully subtracting the continua, and the extended sizes were defined by the area with a surface brightness larger than ${\rm 3\times 10^{-15} erg\,s^{-1}cm^{-2}arcsec^{-2}}$. 
They found a similar slope as \citet{Liu2014} but with a large scatter.
\citet{Fischer2018} updated the results of \citet{Schmitt2003b} with 12 new QSOs and obtained a slope of 0.42, which is close to our results.
In addition, our results are compatible with \citet{Husemann2014} and \citet{Bae2017} who used the flux weighted size. 

At the high luminosity end of the size-luminosity relation, a flattening in slope has been observed \citep{Netzer2004, Hainline2013, Hainline2014, Liu2014}, which is usually interpreted as a maximum size of the ENLR. 
Beyond that radius, the column density may be too low to support [O III] emission \citep{Stern2016} or photons are mostly obscured by the inner clouds \citep{Dempsey2018}.
Since we focused on low luminosity objects with a lower surface brightness cutoff, our results are not sensitive to this upper size limit.

\section{Discussion}
\begin{figure}
  \centering
  \includegraphics[width=0.9\linewidth]{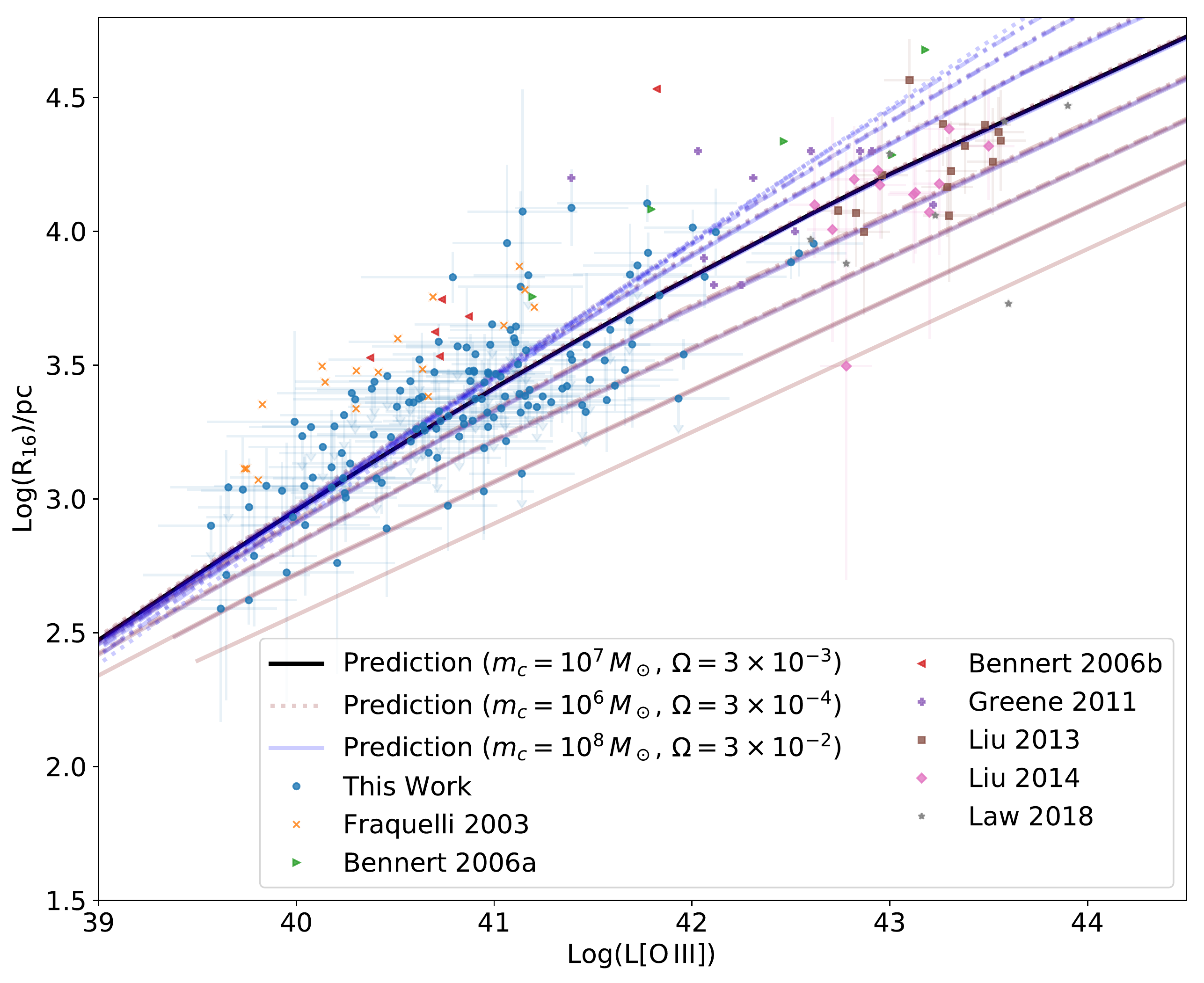}
  \caption{
  Model predictions of \citet{Dempsey2018} are superimposed on our data points. 
  The black line shows the best fitting model, with cloud mass of ${\rm 10^7 M_\odot}$ and a covering factor of ${\rm \Omega = 3\times10^{-3}}$. 
  Other lines are the models with $\pm$1 dex variation in the free parameters. 
  Higher cloud mass ($m_c$) models are shown in blue as clouds mass change from ${\rm 10^6 M_\odot}$ to ${\rm 10^8 M_\odot}$ with colors change from orange to blue.
  Higher covering factor ($\Omega$) models are shown in thicker lines as covering factor changes from $3\times10^{-4}$ to $3\times10^{-2}$ with lines change from dotted to solid.
  }
  \label{fig:model}
\end{figure}

In the standard ionization model, the dimensionless ionization parameter is defined as the ratio between the ionizing photon density and electron density: $U = Q(H_0)/4\pi r^2 c n_e$, where $Q(H_0)$ is the recombination number of hydrogen which is equal to ionizing photons for gas cloud at photoionization equilibrium \citep{Osterbrock2006_Book}.
Based on the assumption that the ionization parameter and electron density both remain constant along the ENLR, $Q$ is proportional to luminosity, which results in $r\propto L^{0.5}$ \citep{Bennert2002, Liu2013, Hainline2013, Liu2014}.
The scenario has been questioned before as some observations appeared to be in conflict with it \citep{Schmitt2003, Greene2011, Liu2013}, but our result over a large dynamic range of the AGN luminosity do not rule out the homogeneous gas model.
Recently, \citet{Dempsey2018} proposed a more detailed model of the ENLR. 
They modeled the (E)NLR as a collection of clouds in pressure equilibrium with the ionizing radiation and the emission line strength like \OIII{} is calculated by Cloudy \citep{Ferland1998, Stern2014}.
By assuming a cloud distribution $n_c \propto r^{-2}$, they predicted a slope of 0.45 for the size-luminosity of ENLR when using a surface brightness cut of ${\rm 10^{-15} erg\,s^{-1}cm^{-2}arcsec^{-2}}$.
If the surface brightness cut at $10^{-16}$\SB{} is adapted, the model fits our data well as shown in Fig. \ref{fig:model}. 
The model takes the masses of clouds ($m_c$) and their covering factor ($\Omega$) as free parameters. 
A cloud mass of $10^7 M_\odot$ with a covering factor of $3\times10^{-3}$ fits our data well as the black solid line in Fig. \ref{fig:all_fitting2}.
Thus, our data also supports the hypothesis that the ENLR consists of a population of photoionized gas clouds which are sufficiently rarefied to be easily ionized by the central AGN source.

It is widely believed that outflows are prevalent in AGN \citep{Harrison2014, Woo2016}, and that they can either clear out the ambient gas in the host or compress it at larger distance. 
Lots of debates have focused on whether or not AGN outflows can suppress star formation of host galaxies \citep{Shi2009, Zhang2016, Zubovas2017, Harrison2017, Harrison2018, Bing2018, Gallagher2019}, even for the low luminous AGN \citep{Cheung2016, Penny2018}. 
AGN outflows may also affect the extension of the ENLRs by changing the distribution of the gas and dust \citep{Liu2014, Dempsey2018}.
However, there may be a significant difference between the extent of the outflow and the extent of the ENLR. 
Based on the decomposition of \OIII{} kinematics, the non-gravitational outflow can be spatially separated in some luminous Seyfert galaxies, and most of them are smaller than the ENLR \citep{Karouzos2016a, Karouzos2016b, Kang2018, Fischer2018}. 
Since the effective radius of an outflow is limited by its energy or momentum, it typically may not reach the size of the ENLR.

Radio-loud AGN can launch powerful jets that survive as conical structures when the jets interact with gas in the ENLR \citep{Baum1992, Humphrey2006} and the synchrotron radiation produced by jets make them luminous in the radio band \citep{Sikora2007, Netzer2013}.
An ENLR caused by this effect was the first to be observed since they tend to be highly extended \citep{Boroson1985, Unger1987}.To pick out the radio-loud galaxies, We match our AGN candidates with the Faint Images of the Radio Sky at Twenty-cm (FIRST) survey \citep{Becker1995} to get the radio luminosity at 22cm. 
Following the typical power-law spectral energy distribution with $\alpha = 0.5$, we extrapolated to obtain the $L_v{\rm (5\,GHz)}$.
The radio-loud AGN who most likely possess relatively strong jets are selected by the ``radio loudness'' R = $L_v{\rm (5\,GHz)} / L_v{\rm (4400\text{\AA})} > 10$ \citep{Netzer2013}.
We only found nine radio-loud AGN and seven of them show a compact radio core from the FIRST radio image.
Even though 5/9 galaxies show asymmetrical ENLRs, all of them follow the overall log-linear relation between the luminosity and size of ENLRs.
Both the number of radio-loud AGN in our sample and their ENLR sizes
These results suggest that the jets may not be the main trigger of ENLR thus have limited influence on the global size-luminosity relation of ENLR.

\section{Conclusions}
In this work, an IFU-based emission line diagnostics is used to identify the ENLRs and a sample of 152 AGN candidates have been found in MaNGA internal data release of MPL-8 with the ENLR features.
Most of the candidates are low-luminosity Seyferts but span about three orders of magnitude in the AGN \OIII{} luminosity.
The key points of this work are listed as follows:
\begin{enumerate}
  \item Based on a surface brightness cut of $10^{-16}$\SB{}, we confirmed the log-linear relation between size and luminosity of ENLR over four orders of magnitude of the AGN \OIII{} luminosity including high luminous quasars . The slope derived by the IFU data is $0.42\pm0.02$, 
    which is more consistent with the model that the ENLR is filled by a population of photoionized clouds \citep{Dempsey2018}.
  \item Kinematic feedbacks have limited influence on the global size-luminosity relation of ENLR in the low luminosity range, as the regions detected outflow mostly smaller than the ENLRs and the radio-loud AGN do not show a more extended ENLRs.
\end{enumerate}

\section*{Acknowledgements}
The authors thank Xiaotong Guo for great help for the SED fitting.
This work is supported by the National Key R\&D Program of China (No. 2018YFA0404502, No. 2017YFA0402704), the National Natural Science Foundation of China (NSFC grants 11825302,
11733002 and 11773013) and the Excellent Youth Foundation of the Jiangsu Scientific Committee (BK20150014).

SBR and RAR acknowledge partial financial support by Funda\c c\~ao de Amparo \`a pesquisa do Estado do RS (FAPERGS). RR thanks CNPq, FAPERGS and CAPES for partial financial supporting this project.

Funding for the Sloan Digital Sky Survey IV has been provided by the Alfred P. Sloan Foundation, the U.S. Department of Energy Office of Science, and the Participating Institutions. SDSS acknowledges support and resources from the Center for High-Performance Computing at the University of Utah. The SDSS web site is www.sdss.org.
SDSS is managed by the Astrophysical Research Consortium for the Participating Institutions of the SDSS Collaboration including the Brazilian Participation Group, the Carnegie Institution for Science, Carnegie Mellon University, the Chilean Participation Group, the French Participation Group, Harvard-Smithsonian Center for Astrophysics, Instituto de Astrofísica de Canarias, The Johns Hopkins University, Kavli Institute for the Physics and Mathematics of the Universe (IPMU) / University of Tokyo, the Korean Participation Group, Lawrence Berkeley National Laboratory, Leibniz Institut für Astrophysik Potsdam (AIP), Max-Planck-Institut für Astronomie (MPIA Heidelberg), Max-Planck-Institut für Astrophysik (MPA Garching), Max-Planck-Institut für Extraterrestrische Physik (MPE), National Astronomical Observatories of China, New Mexico State University, New York University, University of Notre Dame, Observatório Nacional / MCTI, The Ohio State University, Pennsylvania State University, Shanghai Astronomical Observatory, United Kingdom Participation Group, Universidad Nacional Autónoma de México, University of Arizona, University of Colorado Boulder, University of Oxford, University of Portsmouth, University of Utah, University of Virginia, University of Washington, University of Wisconsin, Vanderbilt University, and Yale University.

This research made use of:
\href{http://www.numpy.org/}{\texttt{NumPy}}, 
    a fundamental package for scientific computing with Python (\citealt{NumPy});
\href{http://www.scipy.org/}{\texttt{SciPy}},
    an open source scientific tool for Python (\citealt{SciPy});
\href{http://matplotlib.org/}{\texttt{Matplotlib}}, 
    a 2-D plotting library for Python (\citealt{Matplotlib});
\href{http://www.astropy.org/}{\texttt{Astropy}}, a community-developed 
    core Python package for astronomy (\citealt{AstroPy}); 
\href{https://ipython.org}{\texttt{IPython}}, 
    an interactive computing system for Python (\citealt{IPython});
\href{https://photutils.readthedocs.io/en/v0.6}{\texttt{photutils}},
    an affiliated package of \texttt{AstroPy} to provide tools for detecting and performing photometry of astronomical sources (\citealt{photutils});
\href{http://www-astro.physics.ox.ac.uk/~mxc/software/#ppxf}{\texttt{pPXF}}
    a software using a maximum penalized likelihood approach to extract the stellar kinematics or stellar population from absorption-line spectra of galaxies (\citealt{Cappellari2004, Cappellari2017});
\href{http://www-astro.physics.ox.ac.uk/~mxc/software/#binning<Paste>}{\texttt{VorBin}}
    a package adaptively spatial bin two-dimensional data to a constant signal-to-noise ratio per bin (\citealt{Cappellari2003});
\href{https://dr15.sdss.org/marvin}{\texttt{Marvin}}
    a toolkit for streamlined access and visualization of the SDSS-IV MaNGA data set (\citealt{Cherinka2018}).

\bibliography{ENLR}

\begin{thebibliography}{}
\makeatletter
\relax
\def\mn@urlcharsother{\let\do\@makeother \do\$\do\&\do\#\do\^\do\_\do\%\do\~}
\def\mn@doi{\begingroup\mn@urlcharsother \@ifnextchar [ {\mn@doi@}
  {\mn@doi@[]}}
\def\mn@doi@[#1]#2{\def\@tempa{#1}\ifx\@tempa\@empty \href
  {http://dx.doi.org/#2} {doi:#2}\else \href {http://dx.doi.org/#2} {#1}\fi
  \endgroup}
\def\mn@eprint#1#2{\mn@eprint@#1:#2::\@nil}
\def\mn@eprint@arXiv#1{\href {http://arxiv.org/abs/#1} {{\tt arXiv:#1}}}
\def\mn@eprint@dblp#1{\href {http://dblp.uni-trier.de/rec/bibtex/#1.xml}
  {dblp:#1}}
\def\mn@eprint@#1:#2:#3:#4\@nil{\def\@tempa {#1}\def\@tempb {#2}\def\@tempc
  {#3}\ifx \@tempc \@empty \let \@tempc \@tempb \let \@tempb \@tempa \fi \ifx
  \@tempb \@empty \def\@tempb {arXiv}\fi \@ifundefined
  {mn@eprint@\@tempb}{\@tempb:\@tempc}{\expandafter \expandafter \csname
  mn@eprint@\@tempb\endcsname \expandafter{\@tempc}}}

\bibitem[\protect\citeauthoryear{{Aguado} et~al.,}{{Aguado}
  et~al.}{2019}]{Aguado2019}
{Aguado} D.~S.,  et~al., 2019, \mn@doi [\apjs] {10.3847/1538-4365/aaf651},
  \href {http://ads.bao.ac.cn/abs/2019ApJS..240...23A} {240, 23}

\bibitem[\protect\citeauthoryear{{Antonucci}}{{Antonucci}}{1993}]{Antonucci1993}
{Antonucci} R.,  1993, \mn@doi [\araa] {10.1146/annurev.aa.31.090193.002353},
  \href {http://adsabs.harvard.edu/abs/1993ARA%26A..31..473A} {31, 473}

\bibitem[\protect\citeauthoryear{{Bae}, {Woo}, {Karouzos}, {Gallo}, {Flohic},
  {Shen}  \& {Yoon}}{{Bae} et~al.}{2017}]{Bae2017}
{Bae} H.-J.,  {Woo} J.-H.,  {Karouzos} M.,  {Gallo} E.,  {Flohic} H.,  {Shen}
  Y.,   {Yoon} S.-J.,  2017, \mn@doi [\apj] {10.3847/1538-4357/aa5f5c}, \href
  {http://adsabs.harvard.edu/abs/2017ApJ...837...91B} {837, 91}

\bibitem[\protect\citeauthoryear{{Baldwin}, {Phillips}  \&
  {Terlevich}}{{Baldwin} et~al.}{1981}]{Baldwin1981}
{Baldwin} J.~A.,  {Phillips} M.~M.,   {Terlevich} R.,  1981, \mn@doi [\pasp]
  {10.1086/130766}, \href {http://adsabs.harvard.edu/abs/1981PASP...93....5B}
  {93, 5}

\bibitem[\protect\citeauthoryear{{Baum}, {Heckman}  \& {van Breugel}}{{Baum}
  et~al.}{1992}]{Baum1992}
{Baum} S.~A.,  {Heckman} T.~M.,   {van Breugel} W.,  1992, \mn@doi [\apj]
  {10.1086/171198}, \href
  {https://ui.adsabs.harvard.edu/abs/1992ApJ...389..208B} {389, 208}

\bibitem[\protect\citeauthoryear{{Becker}, {White}  \& {Helfand}}{{Becker}
  et~al.}{1995}]{Becker1995}
{Becker} R.~H.,  {White} R.~L.,   {Helfand} D.~J.,  1995, \mn@doi [\apj]
  {10.1086/176166}, \href {http://adsabs.harvard.edu/abs/1995ApJ...450..559B}
  {450, 559}

\bibitem[\protect\citeauthoryear{{Belfiore} et~al.,}{{Belfiore}
  et~al.}{2016}]{Belfiore2016}
{Belfiore} F.,  et~al., 2016, \mn@doi [\mnras] {10.1093/mnras/stw1234}, \href
  {http://ads.bao.ac.cn/abs/2016MNRAS.461.3111B} {461, 3111}

\bibitem[\protect\citeauthoryear{{Belfiore} et~al.,}{{Belfiore}
  et~al.}{2019}]{Belfiore2019}
{Belfiore} F.,  et~al., 2019, arXiv e-prints, \href
  {http://ads.bao.ac.cn/abs/2019arXiv190100866B} {}

\bibitem[\protect\citeauthoryear{{Bennert}, {Falcke}, {Schulz}, {Wilson}  \&
  {Wills}}{{Bennert} et~al.}{2002}]{Bennert2002}
{Bennert} N.,  {Falcke} H.,  {Schulz} H.,  {Wilson} A.~S.,   {Wills} B.~J.,
  2002, \mn@doi [\apjl] {10.1086/342420}, \href
  {http://adsabs.harvard.edu/abs/2002ApJ...574L.105B} {574, L105}

\bibitem[\protect\citeauthoryear{{Bennert}, {Jungwiert}, {Komossa}, {Haas}  \&
  {Chini}}{{Bennert} et~al.}{2006a}]{Bennert2006a}
{Bennert} N.,  {Jungwiert} B.,  {Komossa} S.,  {Haas} M.,   {Chini} R.,  2006a,
  \mn@doi [\aap] {10.1051/0004-6361:20065319}, \href
  {http://adsabs.harvard.edu/abs/2006A%26A...456..953B} {456, 953}

\bibitem[\protect\citeauthoryear{{Bennert}, {Jungwiert}, {Komossa}, {Haas}  \&
  {Chini}}{{Bennert} et~al.}{2006b}]{Bennert2006b}
{Bennert} N.,  {Jungwiert} B.,  {Komossa} S.,  {Haas} M.,   {Chini} R.,  2006b,
  \mn@doi [\aap] {10.1051/0004-6361:20065477}, \href
  {http://adsabs.harvard.edu/abs/2006A%26A...459...55B} {459, 55}

\bibitem[\protect\citeauthoryear{{Bing} et~al.,}{{Bing}
  et~al.}{2018}]{Bing2018}
{Bing} L.,  et~al., 2018, \mn@doi [\mnras] {10.1093/mnras/sty2662}, \href
  {http://adsabs.harvard.edu/abs/2018MNRAS.tmp.2546B} {}

\bibitem[\protect\citeauthoryear{{Blanton} et~al.,}{{Blanton}
  et~al.}{2017}]{Blanton2017}
{Blanton} M.~R.,  et~al., 2017, \mn@doi [\aj] {10.3847/1538-3881/aa7567}, \href
  {http://adsabs.harvard.edu/abs/2017AJ....154...28B} {154, 28}

\bibitem[\protect\citeauthoryear{{Boroson}, {Persson}  \& {Oke}}{{Boroson}
  et~al.}{1985}]{Boroson1985}
{Boroson} T.~A.,  {Persson} S.~E.,   {Oke} J.~B.,  1985, \mn@doi [\apj]
  {10.1086/163219}, \href {http://adsabs.harvard.edu/abs/1985ApJ...293..120B}
  {293, 120}

\bibitem[\protect\citeauthoryear{{Bradley}, {Kaiser}  \& {Baan}}{{Bradley}
  et~al.}{2004}]{Bradley2004}
{Bradley} L.~D.,  {Kaiser} M.~E.,   {Baan} W.~A.,  2004, \mn@doi [\apj]
  {10.1086/381680}, \href {http://adsabs.harvard.edu/abs/2004ApJ...603..463B}
  {603, 463}

\bibitem[\protect\citeauthoryear{Bradley et~al.,}{Bradley
  et~al.}{2017}]{photutils}
Bradley L.,  et~al., 2017, astropy/photutils: v0.4,
  \mn@doi{10.5281/zenodo.1039309}, \url
  {https://doi.org/10.5281/zenodo.1039309}

\bibitem[\protect\citeauthoryear{{Bundy} et~al.,}{{Bundy}
  et~al.}{2015}]{Bundy2015}
{Bundy} K.,  et~al., 2015, \mn@doi [\apj] {10.1088/0004-637X/798/1/7}, \href
  {http://adsabs.harvard.edu/abs/2015ApJ...798....7B} {798, 7}

\bibitem[\protect\citeauthoryear{{Calzetti}}{{Calzetti}}{2001}]{Calzetti2001}
{Calzetti} D.,  2001, \mn@doi [\pasp] {10.1086/324269}, \href
  {http://adsabs.harvard.edu/abs/2001PASP..113.1449C} {113, 1449}

\bibitem[\protect\citeauthoryear{{Cappellari}}{{Cappellari}}{2017}]{Cappellari2017}
{Cappellari} M.,  2017, \mn@doi [\mnras] {10.1093/mnras/stw3020}, \href
  {http://ads.bao.ac.cn/abs/2017MNRAS.466..798C} {466, 798}

\bibitem[\protect\citeauthoryear{{Cappellari} \& {Copin}}{{Cappellari} \&
  {Copin}}{2003}]{Cappellari2003}
{Cappellari} M.,  {Copin} Y.,  2003, \mn@doi [\mnras]
  {10.1046/j.1365-8711.2003.06541.x}, \href
  {http://adsabs.harvard.edu/abs/2003MNRAS.342..345C} {342, 345}

\bibitem[\protect\citeauthoryear{{Cappellari} \& {Emsellem}}{{Cappellari} \&
  {Emsellem}}{2004}]{Cappellari2004}
{Cappellari} M.,  {Emsellem} E.,  2004, \mn@doi [\pasp] {10.1086/381875}, \href
  {http://ads.bao.ac.cn/abs/2004PASP..116..138C} {116, 138}

\bibitem[\protect\citeauthoryear{{Cherinka} et~al.,}{{Cherinka}
  et~al.}{2018}]{Cherinka2018}
{Cherinka} B.,  et~al., 2018, arXiv e-prints, \href
  {http://adsabs.harvard.edu/abs/2018arXiv181203833C} {}

\bibitem[\protect\citeauthoryear{{Cheung} et~al.,}{{Cheung}
  et~al.}{2016}]{Cheung2016}
{Cheung} E.,  et~al., 2016, \mn@doi [\nat] {10.1038/nature18006}, \href
  {http://adsabs.harvard.edu/abs/2016Natur.533..504C} {533, 504}

\bibitem[\protect\citeauthoryear{{Cid Fernandes}, {Stasi{\'n}ska},
  {Schlickmann}, {Mateus}, {Vale Asari}, {Schoenell}  \& {Sodr{\'e}}}{{Cid
  Fernandes} et~al.}{2010}]{Cid_Fernandes2010}
{Cid Fernandes} R.,  {Stasi{\'n}ska} G.,  {Schlickmann} M.~S.,  {Mateus} A.,
  {Vale Asari} N.,  {Schoenell} W.,   {Sodr{\'e}} L.,  2010, \mn@doi [\mnras]
  {10.1111/j.1365-2966.2009.16185.x}, \href
  {http://adsabs.harvard.edu/abs/2010MNRAS.403.1036C} {403, 1036}

\bibitem[\protect\citeauthoryear{{Dempsey} \& {Zakamska}}{{Dempsey} \&
  {Zakamska}}{2018}]{Dempsey2018}
{Dempsey} R.,  {Zakamska} N.~L.,  2018, \mn@doi [\mnras]
  {10.1093/mnras/sty941}, \href
  {http://adsabs.harvard.edu/abs/2018MNRAS.477.4615D} {477, 4615}

\bibitem[\protect\citeauthoryear{{Dopita} \& {Sutherland}}{{Dopita} \&
  {Sutherland}}{1995}]{Dopita1995}
{Dopita} M.~A.,  {Sutherland} R.~S.,  1995, \mn@doi [\apj] {10.1086/176596},
  \href {http://adsabs.harvard.edu/abs/1995ApJ...455..468D} {455, 468}

\bibitem[\protect\citeauthoryear{{Dopita} \& {Sutherland}}{{Dopita} \&
  {Sutherland}}{1996}]{Dopita1996}
{Dopita} M.~A.,  {Sutherland} R.~S.,  1996, \mn@doi [\apjs] {10.1086/192255},
  \href {http://adsabs.harvard.edu/abs/1996ApJS..102..161D} {102, 161}

\bibitem[\protect\citeauthoryear{{Dopita}, {Groves}, {Sutherland}, {Binette}
  \& {Cecil}}{{Dopita} et~al.}{2002}]{Dopita2002}
{Dopita} M.~A.,  {Groves} B.~A.,  {Sutherland} R.~S.,  {Binette} L.,   {Cecil}
  G.,  2002, \mn@doi [\apj] {10.1086/340429}, \href
  {http://adsabs.harvard.edu/abs/2002ApJ...572..753D} {572, 753}

\bibitem[\protect\citeauthoryear{{Drory} et~al.,}{{Drory}
  et~al.}{2015}]{Drory2015}
{Drory} N.,  et~al., 2015, \mn@doi [\aj] {10.1088/0004-6256/149/2/77}, \href
  {http://adsabs.harvard.edu/abs/2015AJ....149...77D} {149, 77}

\bibitem[\protect\citeauthoryear{{Fabian}}{{Fabian}}{2012}]{Fabian2012}
{Fabian} A.~C.,  2012, \mn@doi [\araa] {10.1146/annurev-astro-081811-125521},
  \href {http://adsabs.harvard.edu/abs/2012ARA%26A..50..455F} {50, 455}

\bibitem[\protect\citeauthoryear{{Ferland}, {Korista}, {Verner}, {Ferguson},
  {Kingdon}  \& {Verner}}{{Ferland} et~al.}{1998}]{Ferland1998}
{Ferland} G.~J.,  {Korista} K.~T.,  {Verner} D.~A.,  {Ferguson} J.~W.,
  {Kingdon} J.~B.,   {Verner} E.~M.,  1998, \mn@doi [\pasp] {10.1086/316190},
  \href {http://ads.bao.ac.cn/abs/1998PASP..110..761F} {110, 761}

\bibitem[\protect\citeauthoryear{{Fischer} et~al.,}{{Fischer}
  et~al.}{2018}]{Fischer2018}
{Fischer} T.~C.,  et~al., 2018, \mn@doi [\apj] {10.3847/1538-4357/aab03e},
  \href {http://adsabs.harvard.edu/abs/2018ApJ...856..102F} {856, 102}

\bibitem[\protect\citeauthoryear{{Fraquelli}, {Storchi-Bergmann}  \&
  {Levenson}}{{Fraquelli} et~al.}{2003}]{Fraquelli2003}
{Fraquelli} H.~A.,  {Storchi-Bergmann} T.,   {Levenson} N.~A.,  2003, \mn@doi
  [\mnras] {10.1046/j.1365-8711.2003.06397.x}, \href
  {http://adsabs.harvard.edu/abs/2003MNRAS.341..449F} {341, 449}

\bibitem[\protect\citeauthoryear{{Fu} \& {Stockton}}{{Fu} \&
  {Stockton}}{2009}]{Fu2009}
{Fu} H.,  {Stockton} A.,  2009, \mn@doi [\apj] {10.1088/0004-637X/690/1/953},
  \href {http://adsabs.harvard.edu/abs/2009ApJ...690..953F} {690, 953}

\bibitem[\protect\citeauthoryear{{Gallagher}, {Maiolino}, {Belfiore}, {Drory},
  {Riffel}  \& {Riffel}}{{Gallagher} et~al.}{2019}]{Gallagher2019}
{Gallagher} R.,  {Maiolino} R.,  {Belfiore} F.,  {Drory} N.,  {Riffel} R.,
  {Riffel} R.~A.,  2019, \mn@doi [\mnras] {10.1093/mnras/stz564}, \href
  {http://ads.bao.ac.cn/abs/2019MNRAS.485.3409G} {485, 3409}

\bibitem[\protect\citeauthoryear{{Greene}, {Zakamska}, {Ho}  \&
  {Barth}}{{Greene} et~al.}{2011}]{Greene2011}
{Greene} J.~E.,  {Zakamska} N.~L.,  {Ho} L.~C.,   {Barth} A.~J.,  2011, \mn@doi
  [\apj] {10.1088/0004-637X/732/1/9}, \href
  {http://adsabs.harvard.edu/abs/2011ApJ...732....9G} {732, 9}

\bibitem[\protect\citeauthoryear{{Groves}, {Dopita}  \& {Sutherland}}{{Groves}
  et~al.}{2004a}]{Groves2004}
{Groves} B.~A.,  {Dopita} M.~A.,   {Sutherland} R.~S.,  2004a, \mn@doi [\apjs]
  {10.1086/421113}, \href {http://adsabs.harvard.edu/abs/2004ApJS..153....9G}
  {153, 9}

\bibitem[\protect\citeauthoryear{{Groves}, {Dopita}  \& {Sutherland}}{{Groves}
  et~al.}{2004b}]{Groves2004b}
{Groves} B.~A.,  {Dopita} M.~A.,   {Sutherland} R.~S.,  2004b, \mn@doi [\apjs]
  {10.1086/421114}, \href {http://adsabs.harvard.edu/abs/2004ApJS..153...75G}
  {153, 75}

\bibitem[\protect\citeauthoryear{{Gunn} et~al.,}{{Gunn}
  et~al.}{2006}]{Gunn2006}
{Gunn} J.~E.,  et~al., 2006, \mn@doi [\aj] {10.1086/500975}, \href
  {http://adsabs.harvard.edu/abs/2006AJ....131.2332G} {131, 2332}

\bibitem[\protect\citeauthoryear{{Hainline}, {Hickox}, {Greene}, {Myers}  \&
  {Zakamska}}{{Hainline} et~al.}{2013}]{Hainline2013}
{Hainline} K.~N.,  {Hickox} R.,  {Greene} J.~E.,  {Myers} A.~D.,   {Zakamska}
  N.~L.,  2013, \mn@doi [\apj] {10.1088/0004-637X/774/2/145}, \href
  {http://adsabs.harvard.edu/abs/2013ApJ...774..145H} {774, 145}

\bibitem[\protect\citeauthoryear{{Hainline}, {Hickox}, {Greene}, {Myers},
  {Zakamska}, {Liu}  \& {Liu}}{{Hainline} et~al.}{2014}]{Hainline2014}
{Hainline} K.~N.,  {Hickox} R.~C.,  {Greene} J.~E.,  {Myers} A.~D.,  {Zakamska}
  N.~L.,  {Liu} G.,   {Liu} X.,  2014, \mn@doi [\apj]
  {10.1088/0004-637X/787/1/65}, \href
  {http://adsabs.harvard.edu/abs/2014ApJ...787...65H} {787, 65}

\bibitem[\protect\citeauthoryear{{Harrison}}{{Harrison}}{2017}]{Harrison2017}
{Harrison} C.~M.,  2017, \mn@doi [Nature Astronomy] {10.1038/s41550-017-0165},
  \href {http://adsabs.harvard.edu/abs/2017NatAs...1E.165H} {1, 0165}

\bibitem[\protect\citeauthoryear{{Harrison}, {Alexander}, {Mullaney}  \&
  {Swinbank}}{{Harrison} et~al.}{2014}]{Harrison2014}
{Harrison} C.~M.,  {Alexander} D.~M.,  {Mullaney} J.~R.,   {Swinbank} A.~M.,
  2014, \mn@doi [\mnras] {10.1093/mnras/stu515}, \href
  {http://adsabs.harvard.edu/abs/2014MNRAS.441.3306H} {441, 3306}

\bibitem[\protect\citeauthoryear{{Harrison}, {Costa}, {Tadhunter},
  {Fl{\"u}tsch}, {Kakkad}, {Perna}  \& {Vietri}}{{Harrison}
  et~al.}{2018}]{Harrison2018}
{Harrison} C.~M.,  {Costa} T.,  {Tadhunter} C.~N.,  {Fl{\"u}tsch} A.,  {Kakkad}
  D.,  {Perna} M.,   {Vietri} G.,  2018, \mn@doi [Nature Astronomy]
  {10.1038/s41550-018-0403-6}, \href
  {http://adsabs.harvard.edu/abs/2018NatAs...2..198H} {2, 198}

\bibitem[\protect\citeauthoryear{{Heckman} \& {Best}}{{Heckman} \&
  {Best}}{2014}]{Heckman2014}
{Heckman} T.~M.,  {Best} P.~N.,  2014, \mn@doi [\araa]
  {10.1146/annurev-astro-081913-035722}, \href
  {http://adsabs.harvard.edu/abs/2014ARA%26A..52..589H} {52, 589}

\bibitem[\protect\citeauthoryear{{Heckman}, {Miley}, {van Breugel}  \&
  {Butcher}}{{Heckman} et~al.}{1981}]{Heckman1981}
{Heckman} T.~M.,  {Miley} G.~K.,  {van Breugel} W.~J.~M.,   {Butcher} H.~R.,
  1981, \mn@doi [\apj] {10.1086/159050}, \href
  {http://adsabs.harvard.edu/abs/1981ApJ...247..403H} {247, 403}

\bibitem[\protect\citeauthoryear{{Heckman}, {Kauffmann}, {Brinchmann},
  {Charlot}, {Tremonti}  \& {White}}{{Heckman} et~al.}{2004}]{Heckman2004}
{Heckman} T.~M.,  {Kauffmann} G.,  {Brinchmann} J.,  {Charlot} S.,  {Tremonti}
  C.,   {White} S.~D.~M.,  2004, \mn@doi [\apj] {10.1086/422872}, \href
  {http://adsabs.harvard.edu/abs/2004ApJ...613..109H} {613, 109}

\bibitem[\protect\citeauthoryear{{Ho}}{{Ho}}{2008}]{Ho2008}
{Ho} L.~C.,  2008, \mn@doi [\araa] {10.1146/annurev.astro.45.051806.110546},
  \href {http://adsabs.harvard.edu/abs/2008ARA%26A..46..475H} {46, 475}

\bibitem[\protect\citeauthoryear{{Hopkins}, {Hernquist}, {Cox}, {Di Matteo},
  {Robertson}  \& {Springel}}{{Hopkins} et~al.}{2006}]{Hopkins2006}
{Hopkins} P.~F.,  {Hernquist} L.,  {Cox} T.~J.,  {Di Matteo} T.,  {Robertson}
  B.,   {Springel} V.,  2006, \mn@doi [\apjs] {10.1086/499298}, \href
  {http://adsabs.harvard.edu/abs/2006ApJS..163....1H} {163, 1}

\bibitem[\protect\citeauthoryear{{Humphrey}, {Villar-Mart{\'\i}n}, {Fosbury},
  {Vernet}  \& {di Serego Alighieri}}{{Humphrey} et~al.}{2006}]{Humphrey2006}
{Humphrey} A.,  {Villar-Mart{\'\i}n} M.,  {Fosbury} R.,  {Vernet} J.,   {di
  Serego Alighieri} S.,  2006, \mn@doi [\mnras]
  {10.1111/j.1365-2966.2006.10224.x}, \href
  {https://ui.adsabs.harvard.edu/abs/2006MNRAS.369.1103H} {369, 1103}

\bibitem[\protect\citeauthoryear{{Hunter}}{{Hunter}}{2007}]{Matplotlib}
{Hunter} J.~D.,  2007, \mn@doi [Computing in Science Engineering]
  {10.1109/MCSE.2007.55}, 9, 90

\bibitem[\protect\citeauthoryear{{Husemann}, {Wisotzki}, {S{\'a}nchez}  \&
  {Jahnke}}{{Husemann} et~al.}{2008}]{Husemann2008}
{Husemann} B.,  {Wisotzki} L.,  {S{\'a}nchez} S.~F.,   {Jahnke} K.,  2008,
  \mn@doi [\aap] {10.1051/0004-6361:200810276}, \href
  {http://adsabs.harvard.edu/abs/2008A%26A...488..145H} {488, 145}

\bibitem[\protect\citeauthoryear{{Husemann}, {Wisotzki}, {S{\'a}nchez}  \&
  {Jahnke}}{{Husemann} et~al.}{2013}]{Husemann2013}
{Husemann} B.,  {Wisotzki} L.,  {S{\'a}nchez} S.~F.,   {Jahnke} K.,  2013,
  \mn@doi [\aap] {10.1051/0004-6361/201220076}, \href
  {http://adsabs.harvard.edu/abs/2013A%26A...549A..43H} {549, A43}

\bibitem[\protect\citeauthoryear{{Husemann}, {Jahnke}, {S{\'a}nchez},
  {Wisotzki}, {Nugroho}, {Kupko}  \& {Schramm}}{{Husemann}
  et~al.}{2014}]{Husemann2014}
{Husemann} B.,  {Jahnke} K.,  {S{\'a}nchez} S.~F.,  {Wisotzki} L.,  {Nugroho}
  D.,  {Kupko} D.,   {Schramm} M.,  2014, \mn@doi [\mnras]
  {10.1093/mnras/stu1167}, \href
  {http://adsabs.harvard.edu/abs/2014MNRAS.443..755H} {443, 755}

\bibitem[\protect\citeauthoryear{Jones, Oliphant, Peterson  et~al.}{Jones
  et~al.}{01  }]{SciPy}
Jones E.,  Oliphant T.,  Peterson P.,   et~al., 2001--, {SciPy}: Open source
  scientific tools for {Python}, \url {http://www.scipy.org/}

\bibitem[\protect\citeauthoryear{Kang \& Woo}{Kang \& Woo}{2018}]{Kang2018}
Kang D.,  Woo J.-H.,  2018, The Astrophysical Journal, 864, 124

\bibitem[\protect\citeauthoryear{{Karouzos}, {Woo}  \& {Bae}}{{Karouzos}
  et~al.}{2016a}]{Karouzos2016a}
{Karouzos} M.,  {Woo} J.-H.,   {Bae} H.-J.,  2016a, \mn@doi [\apj]
  {10.3847/0004-637X/819/2/148}, \href
  {http://adsabs.harvard.edu/abs/2016ApJ...819..148K} {819, 148}

\bibitem[\protect\citeauthoryear{{Karouzos}, {Woo}  \& {Bae}}{{Karouzos}
  et~al.}{2016b}]{Karouzos2016b}
{Karouzos} M.,  {Woo} J.-H.,   {Bae} H.-J.,  2016b, \mn@doi [\apj]
  {10.3847/1538-4357/833/2/171}, \href
  {http://adsabs.harvard.edu/abs/2016ApJ...833..171K} {833, 171}

\bibitem[\protect\citeauthoryear{{Kauffmann} \& {Heckman}}{{Kauffmann} \&
  {Heckman}}{2009}]{Kauffmann2009}
{Kauffmann} G.,  {Heckman} T.~M.,  2009, \mn@doi [\mnras]
  {10.1111/j.1365-2966.2009.14960.x}, \href
  {http://adsabs.harvard.edu/abs/2009MNRAS.397..135K} {397, 135}

\bibitem[\protect\citeauthoryear{{Kauffmann} et~al.,}{{Kauffmann}
  et~al.}{2003}]{Kauffmann2003}
{Kauffmann} G.,  et~al., 2003, \mn@doi [\mnras]
  {10.1111/j.1365-2966.2003.07154.x}, \href
  {http://adsabs.harvard.edu/abs/2003MNRAS.346.1055K} {346, 1055}

\bibitem[\protect\citeauthoryear{{Keel} et~al.,}{{Keel}
  et~al.}{2012}]{Keel2012}
{Keel} W.~C.,  et~al., 2012, \mn@doi [\mnras]
  {10.1111/j.1365-2966.2011.20101.x}, \href
  {http://adsabs.harvard.edu/abs/2012MNRAS.420..878K} {420, 878}

\bibitem[\protect\citeauthoryear{{Kelly}}{{Kelly}}{2007}]{Kelly2007}
{Kelly} B.~C.,  2007, \mn@doi [\apj] {10.1086/519947}, \href
  {http://ads.bao.ac.cn/abs/2007ApJ...665.1489K} {665, 1489}

\bibitem[\protect\citeauthoryear{{Kewley}, {Dopita}, {Sutherland}, {Heisler}
  \& {Trevena}}{{Kewley} et~al.}{2001}]{Kewley2001}
{Kewley} L.~J.,  {Dopita} M.~A.,  {Sutherland} R.~S.,  {Heisler} C.~A.,
  {Trevena} J.,  2001, \mn@doi [\apj] {10.1086/321545}, \href
  {http://adsabs.harvard.edu/abs/2001ApJ...556..121K} {556, 121}

\bibitem[\protect\citeauthoryear{{Kewley}, {Groves}, {Kauffmann}  \&
  {Heckman}}{{Kewley} et~al.}{2006}]{Kewley2006}
{Kewley} L.~J.,  {Groves} B.,  {Kauffmann} G.,   {Heckman} T.,  2006, \mn@doi
  [\mnras] {10.1111/j.1365-2966.2006.10859.x}, \href
  {http://adsabs.harvard.edu/abs/2006MNRAS.372..961K} {372, 961}

\bibitem[\protect\citeauthoryear{{King} \& {Pounds}}{{King} \&
  {Pounds}}{2015}]{King2015}
{King} A.,  {Pounds} K.,  2015, \mn@doi [\araa]
  {10.1146/annurev-astro-082214-122316}, \href
  {http://adsabs.harvard.edu/abs/2015ARA%26A..53..115K} {53, 115}

\bibitem[\protect\citeauthoryear{{Kormendy} \& {Ho}}{{Kormendy} \&
  {Ho}}{2013}]{Kormendy2013}
{Kormendy} J.,  {Ho} L.~C.,  2013, \mn@doi [\araa]
  {10.1146/annurev-astro-082708-101811}, \href
  {http://adsabs.harvard.edu/abs/2013ARA%26A..51..511K} {51, 511}

\bibitem[\protect\citeauthoryear{{Lacerda} et~al.,}{{Lacerda}
  et~al.}{2018}]{Lacerda2018}
{Lacerda} E.~A.~D.,  et~al., 2018, \mn@doi [\mnras] {10.1093/mnras/stx3022},
  \href {http://adsabs.harvard.edu/abs/2018MNRAS.474.3727L} {474, 3727}

\bibitem[\protect\citeauthoryear{{Law} et~al.,}{{Law} et~al.}{2015}]{Law2015}
{Law} D.~R.,  et~al., 2015, \mn@doi [\aj] {10.1088/0004-6256/150/1/19}, \href
  {http://adsabs.harvard.edu/abs/2015AJ....150...19L} {150, 19}

\bibitem[\protect\citeauthoryear{{Law} et~al.,}{{Law} et~al.}{2016}]{Law2016}
{Law} D.~R.,  et~al., 2016, \mn@doi [\aj] {10.3847/0004-6256/152/4/83}, \href
  {http://ads.bao.ac.cn/abs/2016AJ....152...83L} {152, 83}

\bibitem[\protect\citeauthoryear{{Law}, {Steidel}, {Chen}, {Strom}, {Rudie}  \&
  {Trainor}}{{Law} et~al.}{2018}]{Law2018}
{Law} D.~R.,  {Steidel} C.~C.,  {Chen} Y.,  {Strom} A.~L.,  {Rudie} G.~C.,
  {Trainor} R.~F.,  2018, \mn@doi [\apj] {10.3847/1538-4357/aae156}, \href
  {https://ui.adsabs.harvard.edu/abs/2018ApJ...866..119L} {866, 119}

\bibitem[\protect\citeauthoryear{{Liu}, {Zakamska}, {Greene}, {Nesvadba}  \&
  {Liu}}{{Liu} et~al.}{2013}]{Liu2013}
{Liu} G.,  {Zakamska} N.~L.,  {Greene} J.~E.,  {Nesvadba} N.~P.~H.,   {Liu} X.,
   2013, \mn@doi [\mnras] {10.1093/mnras/stt051}, \href
  {http://adsabs.harvard.edu/abs/2013MNRAS.430.2327L} {430, 2327}

\bibitem[\protect\citeauthoryear{{Liu}, {Zakamska}  \& {Greene}}{{Liu}
  et~al.}{2014}]{Liu2014}
{Liu} G.,  {Zakamska} N.~L.,   {Greene} J.~E.,  2014, \mn@doi [\mnras]
  {10.1093/mnras/stu974}, \href
  {http://adsabs.harvard.edu/abs/2014MNRAS.442.1303L} {442, 1303}

\bibitem[\protect\citeauthoryear{{McCarthy}, {van Breugel}, {Spinrad}  \&
  {Djorgovski}}{{McCarthy} et~al.}{1987}]{McCarthy1987}
{McCarthy} P.~J.,  {van Breugel} W.,  {Spinrad} H.,   {Djorgovski} S.,  1987,
  \mn@doi [\apjl] {10.1086/185000}, \href
  {http://adsabs.harvard.edu/abs/1987ApJ...321L..29M} {321, L29}

\bibitem[\protect\citeauthoryear{{Nesvadba}, {Lehnert}, {De Breuck}, {Gilbert}
  \& {van Breugel}}{{Nesvadba} et~al.}{2008}]{Nesvadba2008}
{Nesvadba} N.~P.~H.,  {Lehnert} M.~D.,  {De Breuck} C.,  {Gilbert} A.~M.,
  {van Breugel} W.,  2008, \mn@doi [\aap] {10.1051/0004-6361:200810346}, \href
  {http://adsabs.harvard.edu/abs/2008A%26A...491..407N} {491, 407}

\bibitem[\protect\citeauthoryear{{Netzer}}{{Netzer}}{2013}]{Netzer2013}
{Netzer} H.,  2013, {The Physics and Evolution of Active Galactic Nuclei}

\bibitem[\protect\citeauthoryear{{Netzer}, {Shemmer}, {Maiolino}, {Oliva},
  {Croom}, {Corbett}  \& {di Fabrizio}}{{Netzer} et~al.}{2004}]{Netzer2004}
{Netzer} H.,  {Shemmer} O.,  {Maiolino} R.,  {Oliva} E.,  {Croom} S.,
  {Corbett} E.,   {di Fabrizio} L.,  2004, \mn@doi [\apj] {10.1086/423608},
  \href {http://adsabs.harvard.edu/abs/2004ApJ...614..558N} {614, 558}

\bibitem[\protect\citeauthoryear{{Obied}, {Zakamska}, {Wylezalek}  \&
  {Liu}}{{Obied} et~al.}{2016}]{Obied2016}
{Obied} G.,  {Zakamska} N.~L.,  {Wylezalek} D.,   {Liu} G.,  2016, \mn@doi
  [\mnras] {10.1093/mnras/stv2850}, \href
  {http://adsabs.harvard.edu/abs/2016MNRAS.456.2861O} {456, 2861}

\bibitem[\protect\citeauthoryear{{Osterbrock} \& {Ferland}}{{Osterbrock} \&
  {Ferland}}{2006}]{Osterbrock2006_Book}
{Osterbrock} D.~E.,  {Ferland} G.~J.,  2006, {Astrophysics of gaseous nebulae
  and active galactic nuclei}.
University Science Books

\bibitem[\protect\citeauthoryear{{Penny} et~al.,}{{Penny}
  et~al.}{2018}]{Penny2018}
{Penny} S.~J.,  et~al., 2018, \mn@doi [\mnras] {10.1093/mnras/sty202}, \href
  {http://adsabs.harvard.edu/abs/2018MNRAS.476..979P} {476, 979}

\bibitem[\protect\citeauthoryear{{Perez} \& {Granger}}{{Perez} \&
  {Granger}}{2007}]{IPython}
{Perez} F.,  {Granger} B.~E.,  2007, \mn@doi [Computing in Science Engineering]
  {10.1109/MCSE.2007.53}, 9, 21

\bibitem[\protect\citeauthoryear{{Rembold} et~al.,}{{Rembold}
  et~al.}{2017}]{Rembold2017}
{Rembold} S.~B.,  et~al., 2017, \mn@doi [\mnras] {10.1093/mnras/stx2264}, \href
  {http://adsabs.harvard.edu/abs/2017MNRAS.472.4382R} {472, 4382}

\bibitem[\protect\citeauthoryear{{Schawinski}, {Thomas}, {Sarzi}, {Maraston},
  {Kaviraj}, {Joo}, {Yi}  \& {Silk}}{{Schawinski}
  et~al.}{2007}]{Schawinski2007}
{Schawinski} K.,  {Thomas} D.,  {Sarzi} M.,  {Maraston} C.,  {Kaviraj} S.,
  {Joo} S.-J.,  {Yi} S.~K.,   {Silk} J.,  2007, \mn@doi [\mnras]
  {10.1111/j.1365-2966.2007.12487.x}, \href
  {http://adsabs.harvard.edu/abs/2007MNRAS.382.1415S} {382, 1415}

\bibitem[\protect\citeauthoryear{{Schmitt}, {Donley}, {Antonucci}, {Hutchings}
  \& {Kinney}}{{Schmitt} et~al.}{2003a}]{Schmitt2003}
{Schmitt} H.~R.,  {Donley} J.~L.,  {Antonucci} R.~R.~J.,  {Hutchings} J.~B.,
  {Kinney} A.~L.,  2003a, \mn@doi [\apjs] {10.1086/377440}, \href
  {http://adsabs.harvard.edu/abs/2003ApJS..148..327S} {148, 327}

\bibitem[\protect\citeauthoryear{{Schmitt}, {Donley}, {Antonucci}, {Hutchings},
  {Kinney}  \& {Pringle}}{{Schmitt} et~al.}{2003b}]{Schmitt2003b}
{Schmitt} H.~R.,  {Donley} J.~L.,  {Antonucci} R.~R.~J.,  {Hutchings} J.~B.,
  {Kinney} A.~L.,   {Pringle} J.~E.,  2003b, \mn@doi [\apj] {10.1086/381224},
  \href {http://adsabs.harvard.edu/abs/2003ApJ...597..768S} {597, 768}

\bibitem[\protect\citeauthoryear{{Shi}, {Rieke}, {Ogle}, {Jiang}  \&
  {Diamond-Stanic}}{{Shi} et~al.}{2009}]{Shi2009}
{Shi} Y.,  {Rieke} G.~H.,  {Ogle} P.,  {Jiang} L.,   {Diamond-Stanic} A.~M.,
  2009, \mn@doi [\apj] {10.1088/0004-637X/703/1/1107}, \href
  {http://adsabs.harvard.edu/abs/2009ApJ...703.1107S} {703, 1107}

\bibitem[\protect\citeauthoryear{{Sikora}, {Stawarz}  \& {Lasota}}{{Sikora}
  et~al.}{2007}]{Sikora2007}
{Sikora} M.,  {Stawarz} {\L}.,   {Lasota} J.-P.,  2007, \mn@doi [\apj]
  {10.1086/511972}, \href
  {https://ui.adsabs.harvard.edu/abs/2007ApJ...658..815S} {658, 815}

\bibitem[\protect\citeauthoryear{{Smee} et~al.,}{{Smee}
  et~al.}{2013}]{Smee2013}
{Smee} S.~A.,  et~al., 2013, \mn@doi [\aj] {10.1088/0004-6256/146/2/32}, \href
  {http://adsabs.harvard.edu/abs/2013AJ....146...32S} {146, 32}

\bibitem[\protect\citeauthoryear{{Sol{\'o}rzano-I{\~n}arrea}, {Tadhunter}  \&
  {Axon}}{{Sol{\'o}rzano-I{\~n}arrea} et~al.}{2001}]{Solorzano2001}
{Sol{\'o}rzano-I{\~n}arrea} C.,  {Tadhunter} C.~N.,   {Axon} D.~J.,  2001,
  \mn@doi [\mnras] {10.1046/j.1365-8711.2001.04334.x}, \href
  {https://ui.adsabs.harvard.edu/abs/2001MNRAS.323..965S} {323, 965}

\bibitem[\protect\citeauthoryear{{Stern}, {Laor}  \& {Baskin}}{{Stern}
  et~al.}{2014}]{Stern2014}
{Stern} J.,  {Laor} A.,   {Baskin} A.,  2014, \mn@doi [\mnras]
  {10.1093/mnras/stt1843}, \href {http://ads.bao.ac.cn/abs/2014MNRAS.438..901S}
  {438, 901}

\bibitem[\protect\citeauthoryear{{Stern}, {Faucher-Gigu{\`e}re}, {Zakamska}  \&
  {Hennawi}}{{Stern} et~al.}{2016}]{Stern2016}
{Stern} J.,  {Faucher-Gigu{\`e}re} C.-A.,  {Zakamska} N.~L.,   {Hennawi} J.~F.,
   2016, \mn@doi [\apj] {10.3847/0004-637X/819/2/130}, \href
  {http://adsabs.harvard.edu/abs/2016ApJ...819..130S} {819, 130}

\bibitem[\protect\citeauthoryear{{Stockton} \& {MacKenty}}{{Stockton} \&
  {MacKenty}}{1987}]{Stockton1987}
{Stockton} A.,  {MacKenty} J.~W.,  1987, \mn@doi [\apj] {10.1086/165227}, \href
  {http://adsabs.harvard.edu/abs/1987ApJ...316..584S} {316, 584}

\bibitem[\protect\citeauthoryear{{Sun} et~al.,}{{Sun} et~al.}{2018}]{Sun2018}
{Sun} A.-L.,  et~al., 2018, \mn@doi [\mnras] {10.1093/mnras/sty1394}, \href
  {http://adsabs.harvard.edu/abs/2018MNRAS.tmp.1342S} {}

\bibitem[\protect\citeauthoryear{{The Astropy Collaboration} et~al.,}{{The
  Astropy Collaboration} et~al.}{2018}]{AstroPy}
{The Astropy Collaboration} et~al., 2018, preprint, \href
  {http://adsabs.harvard.edu/abs/2018arXiv180102634T} {} (\mn@eprint {arXiv}
  {1801.02634})

\bibitem[\protect\citeauthoryear{{Unger}, {Pedlar}, {Axon}, {Whittle}, {Meurs}
  \& {Ward}}{{Unger} et~al.}{1987}]{Unger1987}
{Unger} S.~W.,  {Pedlar} A.,  {Axon} D.~J.,  {Whittle} M.,  {Meurs} E.~J.~A.,
  {Ward} M.~J.,  1987, \mn@doi [\mnras] {10.1093/mnras/228.3.671}, \href
  {http://adsabs.harvard.edu/abs/1987MNRAS.228..671U} {228, 671}

\bibitem[\protect\citeauthoryear{{Wake} et~al.,}{{Wake}
  et~al.}{2017}]{Wake2017}
{Wake} D.~A.,  et~al., 2017, \mn@doi [\aj] {10.3847/1538-3881/aa7ecc}, \href
  {http://adsabs.harvard.edu/abs/2017AJ....154...86W} {154, 86}

\bibitem[\protect\citeauthoryear{{Westfall} et~al.,}{{Westfall}
  et~al.}{2019}]{Westfall2019}
{Westfall} K.~B.,  et~al., 2019, arXiv e-prints, \href
  {http://ads.bao.ac.cn/abs/2019arXiv190100856W} {}

\bibitem[\protect\citeauthoryear{{Woo}, {Bae}, {Son}  \& {Karouzos}}{{Woo}
  et~al.}{2016}]{Woo2016}
{Woo} J.-H.,  {Bae} H.-J.,  {Son} D.,   {Karouzos} M.,  2016, \mn@doi [\apj]
  {10.3847/0004-637X/817/2/108}, \href
  {http://adsabs.harvard.edu/abs/2016ApJ...817..108W} {817, 108}

\bibitem[\protect\citeauthoryear{{Wright} et~al.,}{{Wright}
  et~al.}{2010}]{Wright2010}
{Wright} E.~L.,  et~al., 2010, \mn@doi [\aj] {10.1088/0004-6256/140/6/1868},
  \href {http://adsabs.harvard.edu/abs/2010AJ....140.1868W} {140, 1868}

\bibitem[\protect\citeauthoryear{{Yan} et~al.,}{{Yan} et~al.}{2016a}]{Yan2016}
{Yan} R.,  et~al., 2016a, \mn@doi [\aj] {10.3847/0004-6256/151/1/8}, \href
  {http://adsabs.harvard.edu/abs/2016AJ....151....8Y} {151, 8}

\bibitem[\protect\citeauthoryear{{Yan} et~al.,}{{Yan} et~al.}{2016b}]{Yan2016b}
{Yan} R.,  et~al., 2016b, \mn@doi [\aj] {10.3847/0004-6256/152/6/197}, \href
  {http://adsabs.harvard.edu/abs/2016AJ....152..197Y} {152, 197}

\bibitem[\protect\citeauthoryear{{Zhang}, {Shi}, {Rieke}, {Xia}, {Wang}, {Sun}
  \& {Wan}}{{Zhang} et~al.}{2016}]{Zhang2016}
{Zhang} Z.,  {Shi} Y.,  {Rieke} G.~H.,  {Xia} X.,  {Wang} Y.,  {Sun} B.,
  {Wan} L.,  2016, \mn@doi [\apjl] {10.3847/2041-8205/819/2/L27}, \href
  {http://adsabs.harvard.edu/abs/2016ApJ...819L..27Z} {819, L27}

\bibitem[\protect\citeauthoryear{{Zhang} et~al.,}{{Zhang}
  et~al.}{2017}]{Zhang2017}
{Zhang} K.,  et~al., 2017, \mn@doi [\mnras] {10.1093/mnras/stw3308}, \href
  {http://adsabs.harvard.edu/abs/2017MNRAS.466.3217Z} {466, 3217}

\bibitem[\protect\citeauthoryear{{Zubovas} \& {Bourne}}{{Zubovas} \&
  {Bourne}}{2017}]{Zubovas2017}
{Zubovas} K.,  {Bourne} M.~A.,  2017, \mn@doi [\mnras] {10.1093/mnras/stx787},
  \href {http://adsabs.harvard.edu/abs/2017MNRAS.468.4956Z} {468, 4956}

\bibitem[\protect\citeauthoryear{{van der Walt}, {Colbert}  \&
  {Varoquaux}}{{van der Walt} et~al.}{2011}]{NumPy}
{van der Walt} S.,  {Colbert} S.~C.,   {Varoquaux} G.,  2011, \mn@doi
  [Computing in Science Engineering] {10.1109/MCSE.2011.37}, 13, 22

\makeatother
\end{thebibliography}
\bibliographystyle{mnras}

\begin{landscape}
\begin{table}  
  \caption{All the AGN candidates with extended narrow line region
  \label{tab:1}}
  \begin{tabular}{lccccccccccc}
    \hline
    plateifu & ra & dec & z & g$_{\rm PSF}$(FWHM) & $\log L_{\rm [O\,III]}$ & $\log L_{\rm 8\mu m}$ & $\log L_{\rm 1.4GHz}$ & R$_{15}$ & R$_{16}$ & resolved & radio loudness \\ 
             & degree & degree & & arcsec & erg/s & erg/s & erg/s & (kpc) & (kpc) & & \\ 
\hline
    7495-1902 & 205.044769 & 26.841041 & 0.0318073 & 2.32 & 39.66$\pm$0.29 & 41.84 & nan & nan & 3.04$\pm$0.06 & False & 0.00 \\
    7815-6104 & 319.193099 & 11.043741 & 0.0806967 & 2.41 & 42.62$\pm$0.25 & 43.60 & 38.64 & 6.51$\pm$0.85 & 9.01$\pm$0.90 & True & 1.81 \\
    7991-6104 & 258.827410 & 57.658770 & 0.0282021 & 2.40 & 40.61$\pm$0.25 & 42.34 & nan & nan & 3.26$\pm$0.23 & False & 0.00 \\
    7991-3702 & 258.158752 & 57.322421 & 0.0266298 & 2.40 & 40.43$\pm$0.25 & 42.39 & nan & 0.43$\pm$0.32 & 1.15$\pm$0.31 & True & 0.00 \\
    8132-6101 & 111.733682 & 41.026691 & 0.129403 & 2.57 & 41.72$\pm$0.28 & 43.83 & nan & nan & 3.87$\pm$0.20 & False & 0.00 \\
    8247-6101 & 136.089598 & 41.481729 & 0.0244671 & 2.72 & 39.93$\pm$0.41 & 42.31 & 37.34 & nan & 1.07$\pm$0.27 & True & 0.33 \\
    8137-3702 & 115.368720 & 44.408794 & 0.131997 & 3.01 & 42.00$\pm$0.25 & 43.90 & nan & 1.38$\pm$1.45 & 10.35$\pm$1.59 & True & 0.00 \\
    8141-1901 & 117.472421 & 45.248483 & 0.0312591 & 2.61 & 40.97$\pm$0.24 & 42.36 & nan & 0.93$\pm$0.36 & 2.94$\pm$0.38 & True & 0.00 \\
    8143-6101 & 121.014201 & 40.802613 & 0.126168 & 2.52 & 42.12$\pm$0.26 & 43.79 & 39.75 & 5.34$\pm$1.35 & 9.94$\pm$3.74 & True & 8.27 \\
    8256-12704 & 166.129408 & 42.624554 & 0.12611 & 2.43 & 41.84$\pm$0.24 & 43.46 & 38.98 & 2.27$\pm$1.41 & 5.77$\pm$1.36 & True & 1.22 \\
    8249-3704 & 137.874763 & 45.468320 & 0.0268253 & 2.52 & 40.88$\pm$0.32 & 42.94 & 37.88 & 0.86$\pm$0.36 & 2.76$\pm$1.12 & True & 2.26 \\
    8319-12705 & 202.128436 & 47.714038 & 0.0607597 & 2.54 & 40.62$\pm$0.46 & 43.07 & nan & nan & 3.52$\pm$0.48 & False & 0.00 \\
    8341-12704 & 189.213253 & 45.651170 & 0.030345 & 2.56 & 40.88$\pm$0.52 & 42.83 & 38.00 & 0.06$\pm$0.19 & 1.35$\pm$0.57 & True & 0.49 \\
    8439-6104 & 143.510355 & 50.027486 & 0.0378229 & 2.47 & 40.77$\pm$0.28 & 42.97 & 38.53 & 0.58$\pm$0.82 & 2.04$\pm$0.83 & True & 1.92 \\
    8452-1901 & 155.885556 & 46.057755 & 0.0257723 & 2.52 & 40.04$\pm$0.25 & 41.66 & nan & nan & 1.12$\pm$0.28 & True & 0.00 \\
    8483-12703 & 245.248314 & 49.001777 & 0.0582143 & 2.42 & 40.20$\pm$46.90 & 42.40 & nan & nan & 1.51$\pm$0.60 & True & 0.00 \\
    8482-12704 & 243.581821 & 50.465611 & 0.0602584 & 2.44 & 40.82$\pm$10.38 & 42.93 & 39.99 & 0.48$\pm$0.62 & 2.14$\pm$0.62 & True & 24.95 \\
    8549-12701 & 240.470871 & 45.351940 & 0.0420468 & 2.36 & 41.13$\pm$0.32 & 42.94 & 38.31 & 1.02$\pm$0.47 & 2.47$\pm$0.47 & True & 1.13 \\
    8465-12704 & 198.141843 & 48.366614 & 0.0558079 & 2.42 & 41.07$\pm$0.28 & 42.63 & 38.32 & 0.44$\pm$0.63 & 9.05$\pm$6.09 & True & 2.06 \\
    8552-12701 & 226.431661 & 44.404902 & 0.0283402 & 2.59 & 40.48$\pm$0.55 & 43.06 & nan & 0.27$\pm$0.46 & 3.20$\pm$2.70 & True & 0.00 \\
    8552-9102 & 229.308914 & 44.018031 & 0.12226 & 2.57 & 40.46$\pm$0.27 & 43.28 & nan & nan & 3.46$\pm$0.41 & False & 0.00 \\
    8318-3704 & 197.891834 & 44.933078 & 0.0247632 & 2.38 & 40.25$\pm$0.27 & 42.31 & 37.78 & 0.37$\pm$0.29 & 1.05$\pm$0.29 & True & 1.41 \\
    8318-6102 & 197.239319 & 45.905447 & 0.12908 & 2.35 & 41.78$\pm$0.27 & 43.51 & 39.87 & 2.43$\pm$1.51 & 8.32$\pm$1.46 & True & 11.83 \\
    8464-6101 & 186.180997 & 44.410771 & 0.125582 & 2.48 & 42.54$\pm$0.24 & 44.14 & 39.62 & 5.23$\pm$1.38 & 8.28$\pm$1.64 & True & 8.35 \\
    8320-3704 & 206.612456 & 22.076742 & 0.0275673 & 2.38 & 40.23$\pm$0.32 & 41.79 & nan & nan & 3.17$\pm$0.18 & False & 0.00 \\
    8550-3704 & 248.426386 & 39.185120 & 0.0298414 & 2.50 & 40.39$\pm$0.29 & 42.23 & 37.48 & nan & 3.24$\pm$0.21 & False & 0.43 \\
    8606-12701 & 255.029870 & 37.839502 & 0.0633343 & 2.58 & 40.95$\pm$0.31 & 43.04 & nan & nan & 1.55$\pm$1.14 & True & 0.00 \\
    8314-3704 & 243.155037 & 39.419024 & 0.0321959 & 2.37 & 39.85$\pm$0.26 & 42.37 & nan & nan & 1.12$\pm$0.36 & True & 0.00 \\
    8611-3704 & 262.996722 & 59.971638 & 0.0291196 & 2.54 & 40.13$\pm$0.38 & 42.22 & 37.42 & nan & 3.19$\pm$0.17 & False & 0.67 \\
    7992-9102 & 254.542084 & 62.415648 & 0.119399 & 2.54 & 41.69$\pm$0.29 & 43.41 & 38.78 & 2.58$\pm$2.02 & 6.90$\pm$3.02 & True & 1.30 \\
    8603-6101 & 247.159333 & 39.551266 & 0.0311758 & 2.62 & 39.76$\pm$0.26 & 42.65 & 39.80 & nan & 0.93$\pm$0.39 & True & 20.48 \\
    8612-12704 & 254.564575 & 39.391464 & 0.0343116 & 2.52 & 41.68$\pm$0.25 & 43.06 & 38.13 & nan & 3.67$\pm$0.37 & False & 1.10 \\
    8588-12704 & 249.557306 & 40.146821 & 0.030363 & 2.51 & 40.71$\pm$0.31 & 42.90 & 37.53 & 0.41$\pm$0.52 & 1.83$\pm$0.93 & True & 0.28 \\
    8602-12701 & 247.048171 & 39.821898 & 0.0267882 & 2.54 & 40.27$\pm$0.29 & 42.37 & 37.24 & 0.27$\pm$0.30 & 1.36$\pm$0.29 & True & 0.13 \\
    8077-6103 & 39.446587 & 0.405085 & 0.0473019 & 2.59 & 40.53$\pm$0.42 & 42.46 & nan & nan & 3.41$\pm$0.31 & False & 0.00 \\
    8147-6102 & 118.627843 & 25.815986 & 0.0631476 & 2.67 & 40.38$\pm$0.39 & 43.11 & nan & nan & 3.41$\pm$0.00 & False & 0.00 \\
    8146-12705 & 118.053214 & 28.772580 & 0.0636542 & 2.46 & 40.18$\pm$0.32 & 42.57 & nan & nan & 1.31$\pm$0.65 & True & 0.00 \\
    8084-6103 & 50.741676 & 0.054137 & 0.035737 & 2.53 & 40.03$\pm$0.38 & 42.68 & 37.68 & nan & 3.24$\pm$0.25 & False & 0.96 \\
    8718-12702 & 120.700706 & 45.034554 & 0.038928 & 2.49 & 41.03$\pm$0.38 & 43.00 & 38.17 & nan & 3.46$\pm$0.44 & False & 0.83 \\
    8718-12701 & 119.182152 & 44.856709 & 0.04992 & 2.45 & 40.87$\pm$0.32 & 42.68 & nan & nan & 3.48$\pm$0.40 & False & 0.00 \\
\hline
    \end{tabular}
  \end{table}
\end{landscape}

\begin{landscape}
\begin{table}  
  \contcaption{}
  \begin{tabular}{lccccccccccc}
    \hline
    plateifu & ra & dec & z & g$_{\rm PSF}$(FWHM) & $\log L_{\rm [O\,III]}$ & $\log L_{\rm 8\mu m}$ & $\log L_{\rm 1.4GHz}$ & R$_{15}$ & R$_{16}$ & resolved & radio loudness \\ 
             & degree & degree & & arcsec & erg/s & erg/s & erg/s & (kpc) & (kpc) & & \\ 
\hline
    8725-9102 & 127.178094 & 45.742555 & 0.049053 & 2.63 & 41.12$\pm$0.34 & 42.97 & nan & nan & 3.50$\pm$0.30 & False & 0.00 \\
    10001-6102 & 132.653992 & 57.359668 & 0.0261046 & 2.59 & 40.86$\pm$0.31 & 42.80 & 37.56 & 0.38$\pm$0.38 & 3.68$\pm$1.87 & True & 0.47 \\
    8715-3702 & 119.920672 & 50.839973 & 0.0543641 & 2.60 & 42.50$\pm$0.22 & 43.96 & 39.66 & 3.46$\pm$0.71 & 7.67$\pm$1.09 & True & 33.51 \\
    8255-6101 & 166.509879 & 43.173473 & 0.0584258 & 2.66 & 40.85$\pm$0.38 & 42.76 & 38.50 & 0.69$\pm$0.77 & 1.91$\pm$0.64 & True & 1.94 \\
    8241-9102 & 127.170800 & 17.581400 & 0.0665263 & 2.44 & 41.13$\pm$0.32 & 43.32 & 38.80 & 0.49$\pm$1.68 & 6.22$\pm$5.11 & True & 3.62 \\
    8241-6102 & 126.059633 & 17.331951 & 0.0372518 & 2.52 & 41.48$\pm$0.45 & 43.78 & 38.65 & nan & 3.45$\pm$0.32 & False & 2.10 \\
    8720-1901 & 121.147928 & 50.708556 & 0.0227214 & 2.43 & 40.51$\pm$0.25 & 41.52 & nan & 0.62$\pm$0.26 & 2.21$\pm$0.29 & True & 0.00 \\
    8547-12701 & 217.629971 & 52.707159 & 0.0448811 & 2.55 & 41.37$\pm$0.30 & 43.43 & 38.46 & 1.61$\pm$0.49 & 2.64$\pm$0.65 & True & 1.17 \\
    8978-12705 & 249.558611 & 41.938810 & 0.0286035 & 2.42 & 39.98$\pm$0.71 & 42.83 & nan & nan & 1.57$\pm$0.75 & True & 0.00 \\
    8978-6102 & 249.371986 & 40.879947 & 0.0263847 & 2.44 & 38.72$\pm$16.91 & 42.61 & 37.44 & nan & nan & True & 0.65 \\
    8978-9101 & 247.907996 & 41.493643 & 0.0303346 & 2.51 & 39.98$\pm$0.51 & 42.45 & nan & nan & 3.18$\pm$0.22 & False & 0.00 \\
    8979-6102 & 241.823389 & 41.403604 & 0.0346392 & 2.28 & 40.04$\pm$0.33 & 42.58 & nan & nan & 0.80$\pm$0.49 & True & 0.00 \\
    8948-12704 & 167.306020 & 49.519432 & 0.0724271 & 2.42 & 41.11$\pm$0.33 & 43.32 & 38.60 & nan & 3.59$\pm$0.28 & False & 1.41 \\
    8946-3701 & 168.957727 & 46.319564 & 0.0532814 & 2.40 & 40.90$\pm$0.26 & 42.97 & nan & nan & 3.48$\pm$0.25 & False & 0.00 \\
    8947-3701 & 168.947800 & 50.401634 & 0.0473068 & 2.34 & 40.97$\pm$0.25 & 43.14 & nan & 0.83$\pm$0.54 & 1.86$\pm$0.59 & True & 0.00 \\
    8945-3703 & 173.911234 & 47.515520 & 0.045503 & 2.66 & 40.26$\pm$1.44 & 42.55 & nan & nan & 1.06$\pm$0.91 & True & 0.00 \\
    8597-3703 & 224.749647 & 48.409855 & 0.0358627 & 2.48 & 40.30$\pm$0.32 & 42.66 & 38.25 & nan & 3.37$\pm$0.44 & False & 0.90 \\
    9026-9101 & 249.318419 & 44.418230 & 0.0314197 & 2.56 & 40.91$\pm$0.29 & 42.82 & 37.88 & nan & 3.37$\pm$0.35 & False & 0.36 \\
    9049-1901 & 247.560973 & 26.206474 & 0.131457 & 2.47 & 42.06$\pm$0.41 & 44.49 & 39.65 & 4.36$\pm$1.36 & 6.77$\pm$1.85 & True & 8.35 \\
    9002-12702 & 222.810069 & 30.692246 & 0.0547198 & 2.47 & 40.25$\pm$nan & 42.83 & nan & nan & 3.50$\pm$0.63 & False & 0.00 \\
    9031-1902 & 241.029075 & 44.549765 & 0.0429606 & 2.68 & 40.72$\pm$0.27 & 42.71 & nan & 0.54$\pm$0.50 & 2.13$\pm$0.52 & True & 0.00 \\
    9027-12704 & 245.346647 & 32.349014 & 0.034659 & 2.38 & 41.13$\pm$0.25 & 42.90 & 37.93 & 0.90$\pm$0.37 & 2.11$\pm$0.37 & True & 0.72 \\
    8982-3703 & 203.190094 & 26.580376 & 0.0470053 & 2.43 & 41.57$\pm$0.24 & 43.03 & 39.17 & 1.21$\pm$0.58 & 2.34$\pm$1.04 & True & 49.44 \\
    7972-6103 & 315.831306 & 10.944169 & 0.0431534 & 2.55 & 40.59$\pm$1.17 & 42.71 & 39.54 & 0.15$\pm$0.43 & 1.69$\pm$0.50 & True & 20.07 \\
    9025-12704 & 246.050764 & 30.162261 & 0.0482447 & 2.46 & 40.62$\pm$0.31 & 42.70 & nan & 0.28$\pm$0.53 & 2.37$\pm$0.59 & True & 0.00 \\
    7958-9101 & 258.495841 & 33.607137 & 0.0386619 & 2.44 & 40.64$\pm$0.29 & 42.66 & 38.11 & 0.58$\pm$0.51 & 2.41$\pm$1.33 & True & 4.74 \\
    9195-3703 & 29.052229 & 14.906639 & 0.026913 & 2.58 & 41.18$\pm$0.33 & 43.04 & nan & nan & 3.41$\pm$0.43 & False & 0.00 \\
    8080-12703 & 49.487801 & -0.169101 & 0.0227967 & 2.60 & 39.99$\pm$0.44 & 42.90 & 38.44 & nan & 1.94$\pm$1.52 & True & 1.42 \\
    9182-6102 & 119.486337 & 39.993365 & 0.0657771 & 2.36 & 41.59$\pm$0.23 & 43.35 & 40.15 & nan & 3.63$\pm$0.44 & False & 66.91 \\
    9193-12701 & 45.954624 & -1.103750 & 0.0136253 & 2.47 & 40.84$\pm$0.23 & 43.34 & 36.93 & 0.83$\pm$0.15 & 2.01$\pm$0.18 & True & 0.16 \\
    8940-12702 & 120.087418 & 26.613527 & 0.0267379 & 2.56 & 41.56$\pm$0.25 & 43.49 & 37.84 & 0.60$\pm$0.40 & 3.30$\pm$0.66 & True & 0.30 \\
    9183-3703 & 121.920806 & 39.004239 & 0.0233453 & 2.73 & 41.04$\pm$0.23 & 43.32 & 38.20 & 0.43$\pm$0.26 & 2.18$\pm$0.27 & True & 1.66 \\
    8993-12705 & 165.391531 & 45.653868 & 0.0294089 & 2.53 & 41.96$\pm$0.30 & 43.38 & 38.40 & 1.71$\pm$0.32 & 3.47$\pm$0.46 & True & 1.02 \\
    8992-3702 & 171.657262 & 51.573041 & 0.0264158 & 2.68 & 40.21$\pm$0.29 & 42.41 & nan & nan & 0.58$\pm$0.55 & True & 0.00 \\
    9485-12705 & 121.779937 & 36.233479 & 0.0323135 & 2.48 & 41.25$\pm$0.33 & 43.88 & nan & nan & 3.38$\pm$1.38 & False & 0.00 \\
    9487-3702 & 123.330544 & 46.147157 & 0.053819 & 2.42 & 40.90$\pm$0.24 & 42.92 & nan & nan & 3.48$\pm$0.30 & False & 0.00 \\
    8989-3703 & 177.440360 & 50.527016 & 0.026442 & 2.45 & 40.77$\pm$0.25 & 42.80 & nan & 0.19$\pm$0.32 & 0.94$\pm$0.37 & True & 0.00 \\
    8984-9102 & 203.850240 & 27.911890 & 0.0268301 & 2.38 & 39.98$\pm$0.44 & 42.29 & nan & nan & 0.86$\pm$0.29 & True & 0.00 \\
    8983-12701 & 203.830208 & 26.424781 & 0.0253738 & 2.58 & 40.46$\pm$0.28 & 42.49 & nan & 0.35$\pm$0.47 & 0.78$\pm$0.46 & True & 0.00 \\
    8311-6104 & 205.282731 & 23.282055 & 0.0263526 & 2.70 & 41.46$\pm$0.33 & 43.48 & 38.87 & nan & 3.33$\pm$0.10 & False & 5.39 \\
\hline
    \end{tabular}
  \end{table}
\end{landscape}

\begin{landscape}
\begin{table}  
  \contcaption{}
  \begin{tabular}{lccccccccccc}
    \hline
    plateifu & ra & dec & z & g$_{\rm PSF}$(FWHM) & $\log L_{\rm [O\,III]}$ & $\log L_{\rm 8\mu m}$ & $\log L_{\rm 1.4GHz}$ & R$_{15}$ & R$_{16}$ & resolved & radio loudness \\ 
             & degree & degree & & arcsec & erg/s & erg/s & erg/s & (kpc) & (kpc) & & \\ 
\hline
    8309-12701 & 208.289083 & 51.812501 & 0.067572 & 2.46 & 40.91$\pm$0.42 & 43.29 & 38.49 & nan & 3.54$\pm$0.33 & False & 1.56 \\
    9507-12704 & 129.600037 & 25.754501 & 0.0181814 & 2.50 & 40.71$\pm$0.32 & 43.43 & 38.99 & nan & 3.15$\pm$0.20 & False & 8.42 \\
    9507-12705 & 129.520694 & 25.329505 & 0.0281762 & 2.50 & 40.67$\pm$0.27 & 42.60 & nan & 0.82$\pm$0.33 & 1.49$\pm$0.36 & True & 0.00 \\
    9508-12704 & 127.105954 & 26.397370 & 0.0808831 & 2.29 & 40.98$\pm$0.43 & 43.47 & 38.73 & nan & 3.58$\pm$0.24 & False & 1.52 \\
    9508-3704 & 127.107818 & 25.014635 & 0.0287366 & 2.38 & 40.73$\pm$0.29 & 42.54 & nan & nan & 3.29$\pm$0.27 & False & 0.00 \\
    9024-12705 & 223.867459 & 32.840028 & 0.0601609 & 2.39 & 40.89$\pm$0.40 & 43.24 & 38.68 & 0.18$\pm$0.48 & 1.96$\pm$1.33 & True & 1.03 \\
    9488-3702 & 126.216411 & 20.991216 & 0.0228872 & 2.36 & 39.62$\pm$0.28 & 41.98 & nan & nan & 0.39$\pm$0.38 & True & 0.00 \\
    9502-9101 & 128.341931 & 25.104925 & 0.0866 & 2.41 & 40.40$\pm$140.78 & 43.11 & 39.11 & nan & 3.65$\pm$1.02 & False & 2.91 \\
    9502-12703 & 129.545574 & 24.895295 & 0.0286559 & 2.46 & 41.93$\pm$0.26 & 43.17 & 39.22 & nan & 3.38$\pm$0.25 & False & 7.59 \\
    9511-12704 & 129.312117 & 4.695876 & 0.0469711 & 2.44 & 40.97$\pm$0.32 & 42.86 & nan & nan & 3.47$\pm$0.40 & False & 0.00 \\
    9511-6104 & 129.362363 & 3.955690 & 0.0470113 & 2.46 & 40.07$\pm$0.31 & 42.45 & 37.79 & nan & 3.27$\pm$0.25 & False & 1.11 \\
    8990-12705 & 173.537567 & 49.254562 & 0.037232 & 2.33 & 41.01$\pm$0.25 & 42.39 & nan & 0.75$\pm$0.43 & 2.93$\pm$0.60 & True & 0.00 \\
    8990-9101 & 173.933151 & 49.037649 & 0.0296172 & 2.37 & 39.65$\pm$0.42 & 42.71 & 38.03 & nan & 0.52$\pm$0.56 & True & 2.18 \\
    8442-9102 & 200.222837 & 32.190761 & 0.0230284 & 2.49 & 40.58$\pm$0.28 & 42.42 & nan & nan & 3.22$\pm$0.23 & False & 0.00 \\
    9048-1902 & 246.255977 & 24.263156 & 0.0503021 & 2.55 & 41.61$\pm$0.23 & 43.12 & 37.99 & 0.40$\pm$0.63 & 2.65$\pm$0.59 & True & 1.02 \\
    9095-12701 & 241.913639 & 23.416866 & 0.0874596 & 2.36 & 40.82$\pm$0.43 & 43.40 & nan & nan & 3.57$\pm$0.02 & False & 0.00 \\
    9196-12703 & 262.399283 & 54.494424 & 0.0818576 & 2.58 & 41.11$\pm$0.34 & 43.69 & 38.66 & nan & 3.64$\pm$0.35 & False & 1.30 \\
    9881-1901 & 204.806913 & 24.893076 & 0.0282062 & 2.53 & 39.79$\pm$0.32 & 41.81 & nan & nan & 0.61$\pm$0.37 & True & 0.00 \\
    9883-12701 & 255.523309 & 31.797429 & 0.0650401 & 2.41 & 40.94$\pm$0.26 & 42.83 & 38.24 & 0.85$\pm$0.81 & 2.37$\pm$0.79 & True & 0.86 \\
    9888-12701 & 235.475823 & 28.133979 & 0.0332197 & 2.56 & 40.70$\pm$0.28 & 42.61 & 38.16 & 0.52$\pm$0.39 & 2.98$\pm$0.52 & True & 0.60 \\
    9893-6102 & 256.196769 & 24.583993 & 0.0425203 & 2.52 & 39.06$\pm$74.36 & 42.54 & nan & nan & nan & True & 0.00 \\
    8656-12705 & 7.387001 & -1.095700 & 0.0585413 & 2.45 & 40.72$\pm$0.46 & 43.45 & 38.77 & nan & 3.87$\pm$0.60 & True & 1.77 \\
    8656-12702 & 7.085553 & -0.217869 & 0.0612581 & 2.49 & 40.28$\pm$0.38 & 42.98 & nan & nan & 3.40$\pm$0.40 & False & 0.00 \\
    9498-3703 & 120.016894 & 23.437849 & 0.0291978 & 2.71 & 40.95$\pm$0.25 & 42.68 & 38.22 & nan & 1.07$\pm$0.45 & True & 2.12 \\
    10216-3704 & 117.239526 & 17.577121 & 0.0286623 & 2.75 & 40.25$\pm$0.26 & 42.37 & 38.15 & 0.34$\pm$0.42 & 1.01$\pm$0.39 & True & 1.61 \\
    10216-12704 & 118.162300 & 18.321606 & 0.0449364 & 2.65 & 41.39$\pm$0.29 & 42.85 & 38.81 & 0.41$\pm$0.56 & 12.25$\pm$4.03 & True & 2.51 \\
    10218-1902 & 119.207582 & 17.380303 & 0.0364051 & 2.34 & 40.57$\pm$0.28 & 42.67 & nan & nan & 3.36$\pm$0.86 & False & 0.00 \\
    10221-6104 & 124.698298 & 24.537430 & 0.024877 & 2.59 & 39.73$\pm$0.28 & 42.11 & nan & nan & 1.09$\pm$0.49 & True & 0.00 \\
    9503-12701 & 119.972832 & 23.390083 & 0.029143 & 2.47 & 41.06$\pm$0.34 & 43.77 & 38.68 & nan & 3.38$\pm$0.38 & False & 2.17 \\
    9503-6102 & 120.996684 & 23.755667 & 0.0293812 & 2.47 & 41.22$\pm$0.27 & 42.66 & 37.75 & nan & 3.34$\pm$0.27 & False & 0.68 \\
    9503-3704 & 121.037072 & 24.558611 & 0.0438319 & 2.41 & 40.19$\pm$0.37 & 42.48 & nan & nan & 3.27$\pm$0.06 & False & 0.00 \\
    9499-12703 & 118.423230 & 26.492699 & 0.0374233 & 2.92 & 40.24$\pm$0.32 & 42.84 & nan & nan & 3.31$\pm$0.24 & False & 0.00 \\
    9499-6104 & 120.082382 & 26.701442 & 0.0278268 & 2.91 & 40.97$\pm$0.30 & 43.07 & 37.86 & 0.99$\pm$0.33 & 2.11$\pm$0.34 & True & 0.66 \\
    9499-6101 & 118.071091 & 25.669081 & 0.0453547 & 2.96 & 40.58$\pm$0.31 & 42.52 & nan & nan & 3.44$\pm$0.40 & False & 0.00 \\
    10215-1902 & 123.857378 & 37.340519 & 0.0397304 & 2.25 & 40.18$\pm$0.30 & 42.27 & nan & nan & 1.10$\pm$0.61 & True & 0.00 \\
    9495-1901 & 122.938744 & 23.473828 & 0.015734 & 2.47 & 39.57$\pm$0.27 & 41.20 & nan & nan & 2.90$\pm$0.09 & False & 0.00 \\
    8723-6104 & 130.407776 & 54.918571 & 0.044562 & 2.29 & 40.65$\pm$41.29 & 43.48 & nan & nan & nan & False & 0.00 \\
    9489-6104 & 126.084373 & 20.533242 & 0.0199314 & 2.42 & 40.41$\pm$0.35 & 42.31 & nan & nan & 3.08$\pm$0.07 & False & 0.00 \\
    8998-12705 & 163.663838 & 47.862282 & 0.0728894 & 2.65 & 41.08$\pm$0.31 & 43.47 & nan & nan & 3.63$\pm$0.20 & False & 0.00 \\
    8981-6104 & 187.419144 & 36.199417 & 0.0717308 & 2.54 & 41.10$\pm$0.34 & 42.90 & nan & nan & 3.60$\pm$0.38 & False & 0.00 \\
\hline
    \end{tabular}
  \end{table}
\end{landscape}

\begin{landscape}
\begin{table}  
  \contcaption{}
  \begin{tabular}{lccccccccccc}
    \hline
    plateifu & ra & dec & z & g$_{\rm PSF}$(FWHM) & $\log L_{\rm [O\,III]}$ & $\log L_{\rm 8\mu m}$ & $\log L_{\rm 1.4GHz}$ & R$_{15}$ & R$_{16}$ & resolved & radio loudness \\ 
             & degree & degree & & arcsec & erg/s & erg/s & erg/s & (kpc) & (kpc) & & \\ 
\hline
    8988-6102 & 186.251194 & 40.157312 & 0.0735606 & 2.51 & 40.99$\pm$0.39 & 43.38 & 39.24 & nan & 3.65$\pm$0.80 & False & 4.35 \\
    10514-12705 & 144.628220 & 2.573280 & 0.0234383 & 2.52 & 40.24$\pm$0.36 & 42.54 & nan & 0.17$\pm$0.26 & 1.20$\pm$0.26 & True & 0.00 \\
    10514-9102 & 145.151611 & 3.577021 & 0.0164514 & 2.44 & 41.14$\pm$0.27 & 42.89 & 37.63 & nan & 3.10$\pm$0.09 & False & 0.40 \\
    10510-12704 & 179.117267 & 55.125209 & 0.00371352 & 2.53 & 39.76$\pm$0.24 & 41.85 & 36.22 & 0.17$\pm$0.04 & 0.42$\pm$0.09 & True & 0.10 \\
    10510-6103 & 177.778667 & 55.078717 & 0.0194702 & 2.57 & 40.82$\pm$0.29 & 42.91 & 38.02 & nan & 3.23$\pm$0.28 & False & 0.59 \\
    10493-3704 & 126.015364 & 51.904331 & 0.0314512 & 2.52 & 41.06$\pm$0.24 & 42.59 & 37.33 & 0.62$\pm$0.35 & 1.65$\pm$0.36 & True & 0.73 \\
    9884-3704 & 206.669999 & 52.476814 & 0.0291777 & 2.47 & 41.17$\pm$0.24 & 42.51 & 37.51 & 1.03$\pm$0.31 & 2.24$\pm$0.34 & True & 0.37 \\
    10494-12701 & 126.358456 & 53.968934 & 0.0641308 & 2.18 & 41.17$\pm$0.28 & 42.99 & nan & nan & 3.84$\pm$1.51 & False & 0.00 \\
    10508-12703 & 184.652029 & 51.414732 & 0.0471993 & 2.72 & 41.29$\pm$0.28 & 43.21 & nan & 1.09$\pm$0.53 & 2.30$\pm$0.69 & True & 0.00 \\
    10508-6101 & 183.371990 & 50.741529 & 0.0307574 & 2.76 & 40.59$\pm$0.25 & 42.87 & 39.48 & 0.19$\pm$0.43 & 2.30$\pm$0.90 & True & 12.15 \\
    10492-12702 & 124.064090 & 57.530542 & 0.0271824 & 2.75 & 40.21$\pm$0.35 & 42.47 & nan & nan & 0.04$\pm$0.06 & True & 0.00 \\
    10492-6103 & 122.972981 & 57.951893 & 0.0277501 & 2.75 & 41.66$\pm$0.24 & 43.33 & 38.48 & 1.69$\pm$0.30 & 3.04$\pm$1.47 & True & 5.81 \\
    9882-9102 & 207.608226 & 23.456934 & 0.0557381 & 2.59 & 40.95$\pm$0.27 & 42.58 & nan & nan & 2.73$\pm$1.14 & True & 0.00 \\
    9882-3701 & 206.626565 & 23.097190 & 0.0299905 & 2.61 & 39.95$\pm$0.34 & 42.44 & 37.44 & nan & 0.53$\pm$0.59 & True & 0.25 \\
    10503-12703 & 160.228870 & 5.991890 & 0.0276125 & 2.47 & 40.79$\pm$0.47 & 43.02 & 38.53 & 0.76$\pm$0.37 & 6.74$\pm$1.50 & True & 3.12 \\
    9091-3704 & 241.944669 & 25.537511 & 0.0406597 & 2.36 & 41.39$\pm$0.28 & 42.92 & 38.08 & 1.90$\pm$0.72 & 3.31$\pm$2.05 & True & 3.42 \\
    9091-12703 & 240.044496 & 27.605140 & 0.03303 & 2.42 & 40.64$\pm$0.42 & 42.47 & nan & nan & 3.27$\pm$0.16 & False & 0.00 \\
    8337-1901 & 214.096447 & 38.190986 & 0.134659 & 2.35 & 41.70$\pm$0.34 & 43.82 & 39.23 & 0.51$\pm$12.60 & 3.79$\pm$2.71 & True & 4.63 \\
    9885-3703 & 241.152724 & 23.663184 & 0.031032 & 2.34 & 40.48$\pm$0.27 & 42.61 & nan & nan & 3.23$\pm$0.09 & False & 0.00 \\
    8334-3703 & 213.230230 & 39.312652 & 0.0250682 & 2.50 & 39.67$\pm$0.60 & 42.12 & nan & nan & 1.60$\pm$1.63 & True & 0.00 \\
    8260-1901 & 182.253728 & 42.475257 & 0.02364 & 2.35 & 40.08$\pm$0.42 & 41.30 & nan & nan & 3.08$\pm$0.10 & False & 0.00 \\
    9892-3703 & 248.383907 & 24.984722 & 0.0594202 & 2.46 & 41.34$\pm$0.28 & 43.04 & nan & 1.05$\pm$0.66 & 2.59$\pm$0.77 & True & 0.00 \\
    8324-6104 & 198.958904 & 46.338830 & 0.0573434 & 2.39 & 41.14$\pm$0.28 & 43.38 & 38.82 & 0.50$\pm$0.87 & 11.87$\pm$12.49 & True & 4.11 \\
    8614-12703 & 258.118529 & 35.884086 & 0.026418 & 2.63 & 41.00$\pm$0.30 & 42.88 & nan & nan & 3.31$\pm$0.24 & False & 0.00 \\
    8614-3703 & 257.001371 & 36.344421 & 0.0362362 & 2.62 & 40.65$\pm$0.26 & 42.77 & 38.48 & 0.37$\pm$0.47 & 1.80$\pm$0.41 & True & 2.40 \\
    9092-1902 & 240.779508 & 24.523990 & 0.0466923 & 2.33 & 41.16$\pm$0.28 & 43.16 & 38.12 & nan & 3.56$\pm$0.62 & False & 1.75 \\
    9032-12701 & 240.475078 & 31.892062 & 0.0449613 & 2.47 & 41.47$\pm$0.32 & 43.11 & nan & 1.13$\pm$1.34 & 3.78$\pm$2.34 & True & 0.00 \\
    9032-12702 & 241.452770 & 30.717059 & 0.0553885 & 2.42 & 41.16$\pm$0.28 & 42.63 & nan & nan & 2.43$\pm$0.69 & True & 0.00 \\
    8593-12705 & 226.937461 & 51.452832 & 0.0459164 & 2.75 & 41.77$\pm$0.45 & 43.34 & 38.31 & 2.80$\pm$0.57 & 12.74$\pm$2.02 & True & 0.74 \\
    9090-9102 & 243.073394 & 28.429551 & 0.0531379 & 2.62 & 40.39$\pm$0.40 & 42.82 & 39.87 & nan & 3.44$\pm$0.46 & False & 25.73 \\
    9090-3701 & 241.717361 & 27.927542 & 0.0460252 & 2.60 & 41.39$\pm$0.27 & 43.57 & 38.29 & nan & 3.54$\pm$0.43 & False & 1.49 \\
    8651-1902 & 313.902141 & -0.636586 & 0.0535281 & 2.48 & 41.45$\pm$0.25 & 43.11 & 39.14 & nan & 3.35$\pm$0.52 & False & 6.71 \\
\hline
    \end{tabular}
  \end{table}
\end{landscape}
\end{document}